%% file: bare1.tex
\documentclass[journal]{IEEEtran}
\ifCLASSINFOpdf
\else
\fi

\usepackage{bm}
\usepackage{graphicx}
\usepackage{subfigure}
\usepackage{epstopdf}
\usepackage{caption}
\usepackage{amsmath}
\usepackage{amssymb}
\usepackage{cases}
\usepackage{algorithm}
\usepackage{algorithmic}
\usepackage{cite}
\usepackage{enumerate}
\usepackage{multirow}
\usepackage{makecell}
\usepackage{color}
\usepackage{romannum}
\usepackage[super]{nth}

\allowdisplaybreaks[1]
\usepackage{tabularx}
\usepackage{subfiles}
\usepackage{titling}

\begin{document}
\title{Molecular and DNA Artificial Neural Networks via Fractional Coding}
%

%\author{Xingyi Liu,~\IEEEmembership{Student Member,~IEEE}; and Keshab K. %Parhi,~\IEEEmembership{Fellow,~IEEE}\\
\author{Xingyi Liu, {\em Student Member, IEEE}; and Keshab K. Parhi, {\em Fellow, IEEE}\\
University of Minnesota\\
Department of Electrical and Computer Engineering\\
Minneapolis, MN, USA}

\maketitle

\begin{abstract}
This paper considers implementation of artificial neural networks (ANNs) using molecular computing and DNA based on {\em fractional coding}. Prior work had addressed molecular two-layer ANNs with binary inputs and arbitrary weights. In prior work using fractional coding, a simple molecular perceptron that computes sigmoid of {\em scaled} weighted sum of the inputs was presented where the inputs and the weights lie between $[-1,1]$. Even for computing the perceptron, the prior approach suffers from two major limitations. First, it cannot compute the sigmoid of the weighted sum, but only the sigmoid of the scaled weighted sum. Second, many machine learning applications require the coefficients to be arbitrarily positive and negative numbers that are not bounded between $[-1,1]$; such numbers cannot be handled by the prior perceptron using fractional coding. This paper makes four contributions. First molecular perceptrons that can handle arbitrary weights and can compute sigmoid of the weighted sums are presented. Thus, these molecular perceptrons are ideal for regression applications and multi-layer ANNs. A new molecular divider is introduced and is used to compute $sigmoid(ax)$ where $a>1$. Second, based on fractional coding, a molecular artificial neural network (ANN) with one hidden layer is presented. Third, a trained ANN classifier with one hidden layer from seizure prediction application from electroencephalogram is mapped to molecular reactions and DNA and their performances are presented. Fourth, molecular activation functions for rectified linear unit (ReLU) and softmax are also presented.
\end{abstract}


% \fi
\IEEEpeerreviewmaketitle

\section{Introduction}

\begin{figure}[htbp]
\centering
\resizebox{0.485\textwidth}{!}{%
\includegraphics{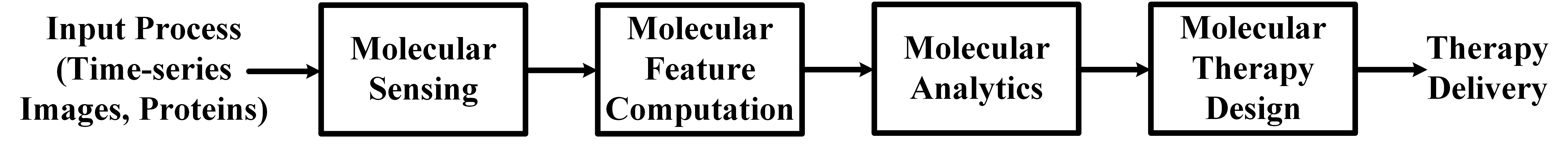}}
\caption{A molecular therapy delivery system.}
\label{fig:therapy}
\end{figure}

Since the pioneering work on DNA computing by Adleman \cite{adleman1994molecular}, there has been growing interest in this field for computing signal processing and machine learning functions. Expected future applications include drug delivery, protein monitoring and molecular controllers. For protein monitoring applications, the goal is to monitor concentration of one or more proteins or rate of growth of these proteins. These proteins may be potential biomarkers for a specific disease. Often the goal may be to monitor spectral content of a certain protein in a specific frequency band~\cite{samoilov2002signal,thurley2014reliable,sumit2015band,park2011seizure}. A drug or therapy can be delivered based on the protein or a molecular biomarker. For example, if a protein concentration or the protein concentration in a specific frequency band exceeds a threshold, a molecular controller can trigger delivery of a therapy. Examples of other features from time series include gene expressions~\cite{ghorbani2018gene} or ratios of band powers in two difference frequency bands \cite{parhi2019discriminative}.  In modern practice, a disease is diagnosed by collecting data from a body sensor, analyzing the data in a computer or laboratory to diagnose a disease, and then delivering a therapy for either prevention or cure. In the proposed molecular biomedicine framework, the sensing, analytics, feature computation and therapy would all be at the same place, i.e., {\em in-vivo}. This paper only addresses design of molecular analytics part; however, this would need to be integrated with molecular sensing and molecular therapy delivery in a complete system (see Fig.~\ref{fig:therapy}).

Several prior publications have addressed molecular implementation of analog systems~\cite{sauro2013synthetic,sarpeshkar1998analog,daniel2013synthetic}. There has also been significant interest in synthetic biology from analog computation circuit perspectives \cite{sarpeshkar2015guest,teo2015synthetic}. However, the focus of this paper is on molecular and DNA implementation of discrete-time systems. These systems can be realized by bimolecular reactions. It has been shown that bimolecular reactions can be mapped to DNA strand displacement (DSD) reactions~\cite{soloveichik2010dna,zhang2009control,yurke2000dna,turberfield2003dna,yurke2003using}. Thus, discrete-time and digital systems can be implemented using DNA.

Simple logic gates, such as AND, OR, NAND, NOR and XOR have been proposed for DNA-based systems~\cite{gardner2000construction,weiss2003genetic,jiang2013digital,jiang2011synchronous,benenson2004autonomous,endy2005foundations,ramalingam2009forward,tamsir2011robust}. There has been significant research on DNA computing for signal processing and digital computing in the past decade~\cite{jiang2012digital,jiang2013discrete,salehi2015molecular,salehi2015markov,salehi2014asynchronous,senum2011rate}. In \cite{jiang2012digital,jiang2013discrete} it was shown that using an asynchronous RGB clock, digital signal processing (DSP) operations such as finite impulse response (FIR) and infinite impulse response (IIR) digital filters can be implemented using DNA. In \cite{jiang2013digital} it was shown that combinational and sequential logic based digital systems can also be implemented by DNA. This was possible by use of bistable reactions~\cite{kim2006construction}. Besides, matrix multiplication and weighted sums can be implemented using combinatorial displacement of DNA~\cite{genot2013combinatorial}.

With ubiquitous interest in machine learning systems and artificial neural networks (ANNs), design and synthesis of molecular machine learning systems and molecular ANNs are naturally of interest.
A phenomenological modeling framework was proposed to explain how the biological regulatory substances act as some neural-network-like computations \textit{in vivo}~\cite{mjolsness1991connectionist}. In~\cite{hjelmfelt1991chemical,blount2017feedforward}, it has been shown that chemical reaction networks with neuron-like properties can be used to construct logic gates. Genetic regulatory networks can be implemented using chemical neural networks~\cite{mestl1996chaos,buchler2003schemes}. DNA Hopfield neural networks~\cite{hopfield1982neural} and DNA multi-layer perceptrons were implemented based on the operations of vector algebra including inner and outer products in DNA vector space by mapping the concentrations of single-stranded DNA to amplitudes of basis vectors~\cite{mills1999article,mills2001experimental}. A DNA transcriptional switch was proposed to design feed-forward networks and winner-take-all networks~\cite{kim2005neural}. A DNA-based molecular pattern classification was developed based on the competitive hybridization reaction between input molecules and a differentially labeled probe mixture~\cite{lim2010vitro}. A multi-gene classification was proposed based on a series of reactions that compute the inner product between RNA inputs and engineered DNA probes~\cite{lopez2018molecular}. In~\cite{poole2017chemical}, Poole \textit{et al.} have shown that a Boltzmann machine could be implemented by using a stochastic chemical reaction network.

Arbitrary linear threshold circuits could be systematically transformed into DNA strand displacement cascades~\cite{qian2011neural,cherry2018scaling}. This implementation allows designers to build complex artificial neural networks like Hopfield associative memory~\cite{qian2011neural}. Winner-take-all neural networks with binary inputs have been implemented using DSD reactions in~\cite{cherry2018scaling}. These molecular and DNA ANNs hold the promise for processing the protein concentrations as features and output a decision variable that can be used to take an action such as monitoring a protein or delivering a drug~\cite{mohammadi2017automated,xie2011multi,li2015modular,miki2015efficient,sayeg2015rationally}. With a variable-gain amplifier~\cite{zhang2010dna,chen2017dna}, these DNA ANNs could also be used to process non-DNA inputs such as RNA or protein~\cite{cherry2018scaling}. 

The DNA neural network in Cherry and Qian~\cite{cherry2018scaling} assumes all inputs to be binary and weights to be non-negative. Furthermore, in Qian, Winfree and Bruck~\cite{qian2011neural}, weights can be arbitrary; however, inputs are still assumed to be binary. The linear classifier with two arbitrary inputs but one positive weight and one negative weight was proposed in Zhang and Seelig~\cite{zhang2010dna}. Furthermore, in Chen and Seelig~\cite{chen2017dna}, the two weights of a linear classifier can be arbitrary; however the summation part of linear classifier is not computed by CRNs~\cite{chen2017dna}. In the current paper, the inputs are bounded between -1 and 1, and the weights can be arbitrary and not necessarily bounded to $[-1,1]$.
% To the best of the authors' knowledge, no prior molecular ANN has been proposed that can handle arbitrary inputs and arbitrary weights that can be either positive or negative.

Using fractional coding, molecular gates for multiplication, addition, inner product, and the perceptron function were presented in \cite{salehi2018computing}. However, the perceptron presented in \cite{salehi2018computing} can only compute $sigmoid(\frac{1}{N}\sum_{i=1}^{N}w_ix_i)$ and cannot compute $sigmoid(\sum_{i=1}^{N}w_ix_i)$. These perceptrons compute sigmoid of the {\em scaled} weighted sum of the inputs; these cannot compute sigmoid of the weighted sum {\em without} scaling. Therefore, these perceptrons are only useful in classification but not regression applications. Furthermore, $w_i$ and $x_i$ must be representable in bipolar format in fractional coding. In many machine learning problems, the weights can be greater than $1$ in magnitude. These weights cannot be handled by the approach in \cite{salehi2018computing}. Thus, two major limitations of the approach in \cite{salehi2018computing} include inability to handle arbitrary weights and inability to compute exact sigmoid function of the weighted sum of the inputs. Whether molecular perceptrons that can implement sigmoid of weighted inputs are synthesizable has remained an open question. This paper shows that such perceptrons can indeed be implemented using molecular computing and DNA where the weights can be arbitrary and only the inputs must correspond to bipolar format. The latter is not a concern as features are typically normalized to the dynamic range $[-1,1]$.

% Since the early pioneering work of \cite{adleman1994molecular}, the field of DNA computing has progressed significantly. For example, several logic
% functions and simple arithmetic operations can be implemented {\em in
% vitro} using bimolecular reactions \cite{zhang2009control,soloveichik2010dna,qian2011neural}. In recent work, fractional coding has been introduced in \cite{salehi2016chemical} for molecular implementation of polynomials using Bernstein expansion. The synergy between fractional coding in molecular computing and stochastic logic in electronic computing was established in \cite{salehi2018computing}. This synergy is significant as it enables every stochastic logic circuit to be translated to a molecular circuit. Based on prior stochastic logic circuit implementations in \cite{liu2016computing}, several complex mathematical functions were implemented
% using molecular computing based on fractional coding \cite{salehi2018computing}. In \cite{salehi2018computing}, a molecular perceptron was implemented using molecular computing; however, the perceptron was only able to perform classification but {\it not} regression. Whether a molecular perceptron can compute a regression has remained an open question so far. This paper shows that a molecular perceptron can not only perform classification but also perform regression.
This paper makes four contributions. First molecular perceptrons that can handle arbitrary weights and can compute sigmoid of the weighted sums are presented. Thus, these molecular perceptrons are ideal for regression applications unlike prior molecular perceptrons \cite{salehi2018computing}. A new molecular divider is introduced and is used to compute sigmoid($ax$) where $a>1$; this circuit overcomes the scaling bottleneck. Second, a molecular implementation of an artificial neural network (ANN) with one hidden layer is presented. It may be noted that such a molecular ANN with non-binary inputs has not been presented before. Third, a trained ANN classifier with one hidden layer from seizure prediction using electroencephalogram~\cite{liu2016machine} is mapped to molecular reactions and DNA. The Software package from David Soloveichik \textit{et al.} is used to simulate the Chemical Reaction Networks and their corresponding DNA implementations presented in this paper~\cite{soloveichik2010dna}.

Although the sigmoid activation function is used for the hidden layer in the application considered in this paper, rectified linear unit (ReLU) and softmax activation functions are used in many neural networks. Molecular and DNA reactions for implementing ReLU and softmax functions are also presented in this paper. We believe molecular ReLU and molecular softmax units have not been presented before.

This paper is organized as follows. Section \Romannum{2} briefly introduces fractional coding. Simple molecular gates using fractional coding are also reviewed in Section \Romannum{2}. Section \Romannum{3} addresses the approaches for mapping specific target functions to molecular reactions. Mapping of molecular reactions to DNA is described in Section \Romannum{4}. Molecular reactions for implementing a single-layer neural network regression (also referred to as a perceptron) and corresponding simulation results with three different sets of weights are described in Section \Romannum{5}. Section \Romannum{6} discusses the molecular and DNA implementations of ANN classifier and presents experimental results of the proposed architectures based on an ANN trained for seizure prediction from electroencephalogram (EEG) signals. Section \Romannum{7} presents molecular and DNA implementations of ReLU and softmax activation functions. Finally, some conclusions are given in Section \Romannum{8}.

% \color{red}
% \section{Prior Work}

% \color{black}

\section{Molecular Gates using Fractional Coding}

{\em Fractional coding} was introduced in ~\cite{salehi2015markov,salehi2016chemical} and was used to realize Markov chains and polynomial computations using molecular reactions and DNA. Fractional coding is inspired by stochastic logic in electronic computing~\cite{gaines1967stochastic,gaines1969stochastic,alaghi2013survey,brown2001stochastic,qian2011transforming,gaudet2003iterative,naderi2011delayed,Bo2015successive,yuan2016belief,tehrani2010majority,liu2015lattice,KK2014architectures,onizawa2015gabor,alaghi2014fast,qian2008synthesis,liu2017computing,liu2016computing,parhi2017analysis,parhi2018stochastic,qian2010architecture}. More recently, stochastic logic was shown to be equivalent to molecular computing based on fractional coding \cite{salehi2018computing}. In fractional coding, each value is encoded using two molecules. For example, value $X$ can be encoded as $X_1$/($X_1~+~X_0$) where $X_1$ and $X_0$, respectively, represent the molecules of type-1 and type-0. In stochastic logic, $X_1$ and $X_0$, respectively, represent the number of $1$ and $0$ bits in a unary bit stream \cite{gaines1967stochastic}. Due to the equivalence, any known stochastic logic system forms the basis for a molecular computing system. It was pointed out in \cite{salehi2018computing} that stochastic digital filters and stochastic error control coders such as low-density parity check~\cite{naderi2011delayed} and polar code decoders~\cite{Bo2015successive,yuan2016belief} can be easily mapped to molecular digital filters and molecular error control coders.

In stochastic logic, the numbers can be represented in either {\em unipolar} or {\em bipolar} format. In the unipolar format, the value of a variable is determined by the fraction of the concentrations of two assigned molecular types:

\begin{equation}
\begin{split}
x=\frac{[X_1]}{[X_0]+[X_1]}\nonumber
\end{split}
\end{equation}

where $[X_0]$ and $[X_1]$ represent the concentrations of the assigned molecular types $X_0$ and $X_1$, respectively. Note that the numbers in the unipolar representation must lie in the unit interval, $[0,1]$.
In the bipolar format, a number $x$ in the range $[-1,1]$ can be represented using molecular types $X_1$ and $X_0$ such that:

\begin{equation}
\begin{split}
x=\frac{[X_1]-[X_0]}{[X_0]+[X_1]}\nonumber
\end{split}
\end{equation}
where $[X_0]$ and $[X_1]$ represent the concentrations of the molecular types $X_0$ and $X_1$, respectively.

% Molecular reactions for the {\tt Mult} and {\tt NMult} units that perform multiplication using unipolar or bipolar representation and {\tt MUX} unit that computes scaled addition have been presented in \cite{salehi2018computing} and are shown in Fig.~\ref{base}.

\begin{figure}[htbp]
\vspace{-1em}
\centering
\resizebox{0.48\textwidth}{!}{%
\includegraphics{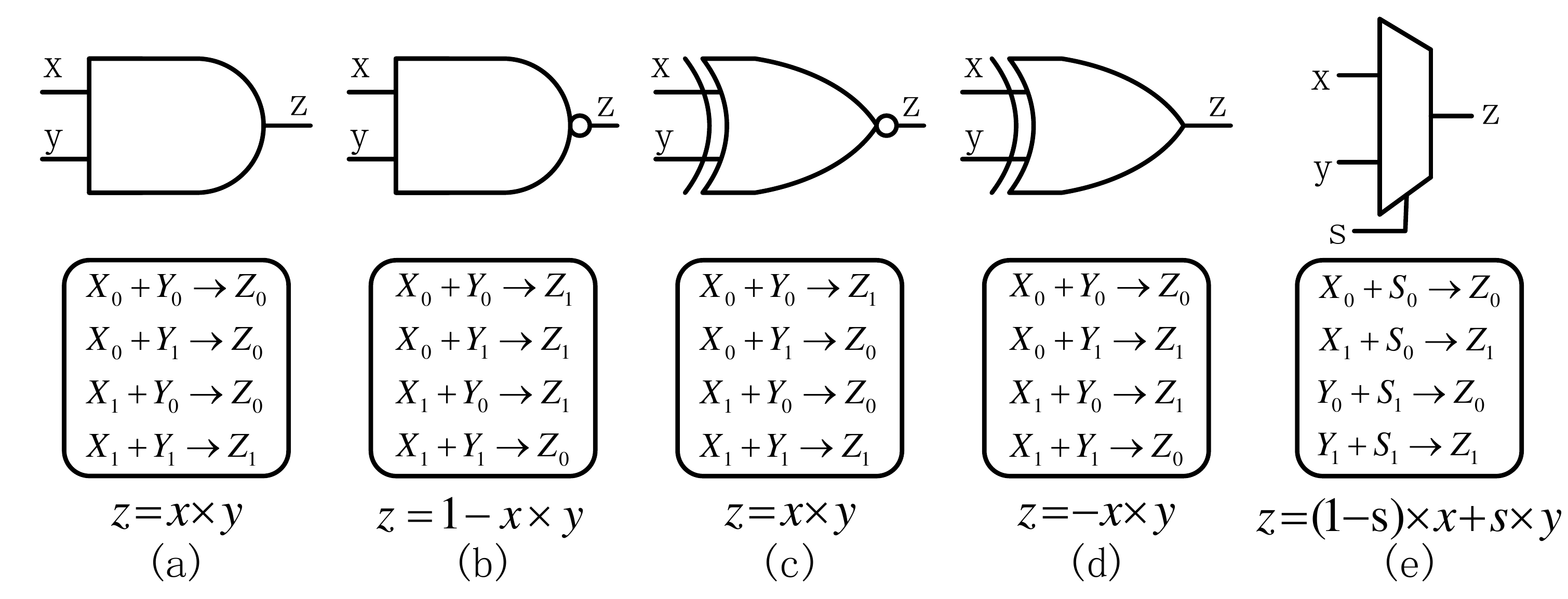}}
\caption{Basic molecular units. (a) Unipolar {\tt Mult} unit. (b) Unipolar {\tt NMult} unit. (c) Bipolar {\tt Mult} unit. (d) Bipolar {\tt NMult} unit. (e) Unipolar/Bipolar {\tt MUX} unit. This is taken from \cite{salehi2018computing}.}
\label{base}
\vspace{-1em}
\end{figure}

Molecular reactions for five basic units have been presented in \cite{salehi2018computing} and are shown in Fig.~\ref{base}. Fig.~\ref{base}(a) shows four molecular reactions of the unipolar {\tt Mult} unit that compute the unipolar product of two unipolar inputs, $z=x\times y$. The unipolar {\tt NMult} unit, consisting of four molecular reactions, shown in Fig.~\ref{base}(b), computes $z=1-x\times y$ where $x$ and $y$ are the unipolar inputs and $z$ is the unipolar output. Bipolar {\tt Mult} and {\tt NMult} units shown in Figs.~\ref{base}(c) and (d) compute $z=x\times y$ and $z=-x\times y$, respectively, where $x$ and $y$ are the bipolar inputs and $z$ is the bipolar output. Unipolar and bipolar {\tt MUX} units that compute $z=(1-s)\times x+s\times y$ can be implemented by using the four molecular reactions shown in Fig.~\ref{base}(e). For unipolar {\tt MUX} unit, $x$, $y$, $z$ and $s$ are all in unipolar format. But for bipolar {\tt MUX} unit, $x$, $y$ and $z$ are in bipolar format whereas $s$ is in unipolar format.

% The four molecular reactions shown in Fig.~\ref{base}(a) compute $z=x\times y$, all in unipolar format; this molecular unit is referred as {\tt Mult} unit. Fig.~\ref{base}(b) shows the molecular reactions that compute $z=1-x\times y$ in unipolar format, referred to as {\tt NMult} unit. Figs.~\ref{base}(c) and (d) illustrate the implementations of \texttt{Mult} unit and \texttt{NMult} unit with bipolar inputs and outputs, respectively. The bipolar \texttt{Mult} unit performs multiplication in the bipolar format. The bipolar \texttt{NMult} unit computes $z=-x\times y$, all in bipolar format. The {\tt MUX} unit that performs scaled addition is shown in Fig.~\ref{base}(e). This unit computes $z=(1-s)\times x+s\times y$. Notice that $x$, $y$ and $z$ can be in the unipolar format or bipolar format, but $s$ must be in unipolar format.

\section{Molecular Reactions for Computing Functions}
This section reviews molecular implementation of exponential functions, and then presents molecular reactions to compute sigmoid, tangent hyperbolic functions and perceptrons. This section reviews implementation of exponential functions as described by Parhi and Liu \cite{parhi2016computing}. Stochastic logic architectures presented in~\cite{brown2001stochastic,qian2010architecture} contain feedback and are not easily adaptable to molecular computing. The tangent hyperbolic and sigmoid functions are inspired by the stochastic logic implementation proposed by Liu and Parhi~\cite{liu2016computing}. The stochastic logic implementation of the sigmoid function in~\cite{li2017neural} does not require explicit computation of the sigmoid function as it makes use of hybrid representation~\cite{parhi2017analysis}; however, it cannot compute $sigmoid(ax)$ where $a>1$ as required for the regression function presented in this paper.

\subsection{Implementation of Exponential Functions in Unipolar Format}
\begin{figure}[htbp]
\vspace{0em}
\centering
\resizebox{0.48\textwidth}{!}{%
\includegraphics{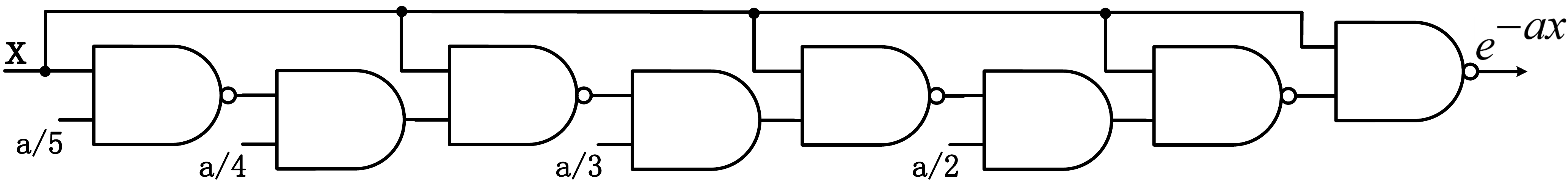}}
\caption{Stochastic implementation of $e^{-ax}$ using \nth{5}-order Maclaurin expansion and Horner's rule \cite{parhi2016computing}.}
\label{enx}
\vspace{-1em}
\end{figure}

The function $e^{-ax}$ $(0<a\leq 1)$ can be approximated as:

\begin{align}
e^{-ax} &\approx 1-ax+\frac{a^2x^2}{2!}-\frac{a^3x^3}{3!}+\frac{a^4x^4}{4!}-\frac{a^5x^5}{5!}\nonumber\\
 &= 1-ax(1-\frac{ax}{2}(1-\frac{ax}{3}(1-\frac{ax}{4}(1-\frac{ax}{5}))))
\label{eqneax}
\end{align}

where $e^{-ax}$ is approximated by a \nth{5}-order truncated Maclaurin series and then reformulated by Horner's rule. In equation (\ref{eqneax}), all coefficients, $a$, $\frac{a}{2}$, $\frac{a}{3}$, $\frac{a}{4}$ and $\frac{a}{5}$, can be represented using unipolar format while $0<a\leq 1$. Fig.~\ref{enx} shows the stochastic implementation of $e^{-ax}$ by cascading \texttt{AND} and \texttt{NAND} gates. The unipolar \texttt{Mult} and unipolar \texttt{NMult} units discussed before compute the same operations as \texttt{AND} and \texttt{NAND} in stochastic implementation, respectively. So equation (\ref{eqneax}) can be implemented using unipolar \texttt{Mult} and unipolar \texttt{NMult} units shown in Figs.~\ref{base} (a) and (b), respectively\cite{salehi2018computing}.

With $a>1$, the coefficients in equation~(\ref{eqneax}) may be larger than $1$ which cannot be represented using fractional representation. Therefore, $e^{-ax}$ $(a>1)$ cannot be directly implemented using this method. However, $e^{-ax}$ $(a>1)$ can be implemented by cascading the output of $e^{-bx}$ ($b\leq 1$)~\cite{parhi2016computing}. For example, the function $e^{-3x}$ can be expressed as:

\begin{equation}
\begin{split}
e^{-3x}=e^{-x}\cdot e^{-x} \cdot e^{-x}.\nonumber
\end{split}
\end{equation}

\begin{figure}[htbp]
\vspace{-0.8em}
\centering
\resizebox{0.3\textwidth}{!}{%
\includegraphics{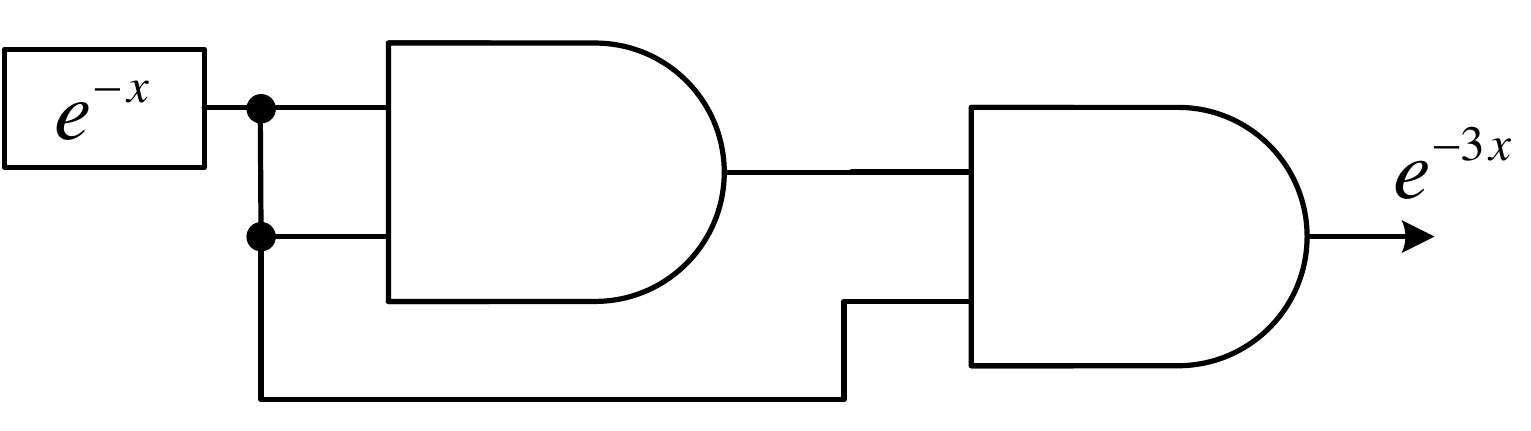}}
\caption{Stochastic implementation of $e^{-3x}$ using $e^{-x}$ and two cascaded \texttt{AND} gates.}
\label{en3x}
\vspace{-0.5em}
\end{figure}

Fig.~\ref{en3x} shows the stochastic implementation of $e^{-3x}$ based on the stochastic circuit of $e^{-x}$ shown in Fig.~\ref{enx}.

\begin{figure}[htbp]
\vspace{-0.8em}
\centering
\resizebox{0.48\textwidth}{!}{%
\includegraphics{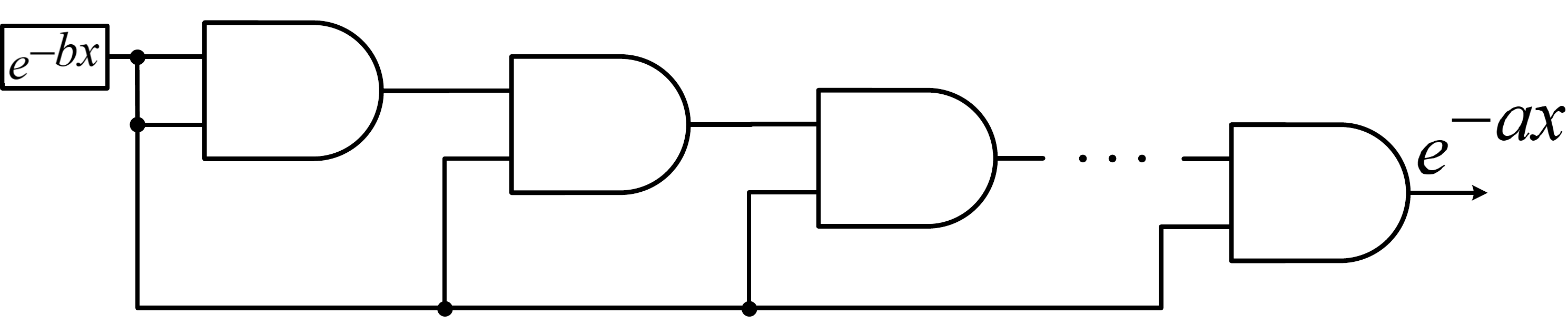}}
\caption{Stochastic implementation of $e^{-ax}$ by cascading $e^{-x}$ and $a-1$ \texttt{AND} gates.}
\label{enax}
\vspace{-1em}
\end{figure}

For any arbitrary $a$ which is greater than $1$, $e^{-ax}$ can be described by:

\begin{equation}
\begin{split}
e^{-ax}=\prod_{n=1}^{N}e^{-bx},b=\frac{a}{N}.\nonumber
\end{split}
\end{equation}

where $0<b\leq 1$ and $N$ is an integer. Notice that, for $b\leq 1$, $e^{-bx}$ can always be implemented using the circuit shown in Fig.~\ref{enx}. Fig.~\ref{enax} shows the implementation of $e^{-ax}$ based on $e^{-bx}$ and $N-1$ cascaded \texttt{AND} gates as presented in \cite{parhi2016computing}. Due to the equivalence of \texttt{AND} in stochastic logic and unipolar \texttt{Mult} unit shown in Fig.~\ref{base}(a), $N-1$ cascaded unipolar \texttt{Mult} units perform the same computation as $N-1$ cascaded \texttt{AND} gates. Therefore, $e^{-ax}$ with large $a$ can be implemented by using unipolar \texttt{Mult} and \texttt{NMult} units as shown in Fig.~\ref{enax}. The \texttt{Mult} and \texttt{NMult} units can be mapped to their equivalent molecular reactions as shown in Fig.~\ref{base}, respectively.

\subsection{Implementation of Division Functions in Unipolar Format}

\begin{figure}[htbp]
\vspace{-1em}
\centering
\resizebox{0.38\textwidth}{!}{
\includegraphics{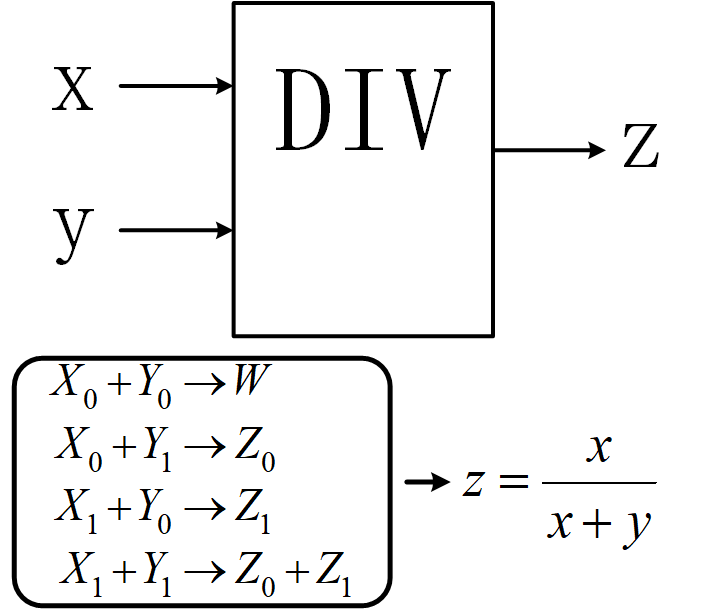}}
\caption{The division unit. This unit calculates $z = \frac{x}{x+y}$, the division of two input variables $x$ and $y$ in unipolar fractional representation. }
\label{div}
\vspace{-1em}
\end{figure}

Implementation of molecular dividers has not been presented before. This section presents a new molecular divider using fractional coding; this design is not inspired by stochastic logic dividers. The four molecular reactions shown in Fig.~\ref{div} compute $z$ as the division of $x$ and $x+y$ using two inputs $x$ and $y$, all in unipolar format. So if $x = \frac{[X1]}{[X0]+[X1]}$ and $y = \frac{[Y1]}{[Y0]+[Y1]}$ then $z = \frac{[Z1]}{[Z0]+[Z1]} = \frac{x}{x+y}$. For these reactions shown in Fig.~\ref{div}, the mass kinetics are described by the ordinary differential equations (ODEs):

\begin{eqnarray}
\frac{d[X_0]}{dt}&=&-[X_0][Y_0]-[X_0][Y_1]\nonumber\\
\frac{d[X_1]}{dt}&=&-[X_1][Y_0]-[X_1][Y_1]\nonumber\\
\frac{d[Y_0]}{dt}&=&-[X_0][Y_0]-[X_1][Y_0]\nonumber\\
\frac{d[Y_1]}{dt}&=&-[X_0][Y_1]-[X_1][Y_1]\nonumber\\
\frac{d[Z_0]}{dt}&=&[X_0][Y_1]+[X_1][Y_1]\nonumber \\
\frac{d[Z_1]}{dt}&=&[X_1][Y_0]+[X_1][Y_1].
\label{divoed}
\end{eqnarray}

Using the ODE equations in (\ref{divoed}) we can prove that the CRN for DIV unit computes the division in fractional coding. The first four equations of (\ref{divoed}) can be rewritten as:

\begin{eqnarray}
\frac{d[X_0]}{X_0}&=&-([Y_0]+[Y_1])dt\nonumber\\
\frac{d[X_1]}{X_1}&=&-([Y_0]+[Y_1])dt\nonumber\\
\frac{d[Y_0]}{Y_0}&=&-([X_0]+[X_1])dt\nonumber\\
\frac{d[Y_1]}{Y_1}&=&-([X_0]+[X_1])dt.
\label{divoed2}
\end{eqnarray}

Comparing the first two equations of (\ref{divoed2}) we have

\begin{eqnarray}
\label{eq:integral}
\int_{0}^{t} \frac{d[X_0]}{[X_0]}=\int_{0}^{t} \frac{d[X_1]}{[X_1]}=\int_{0}^{t}-([Y_0]+[Y_1])dt \nonumber\\
\end{eqnarray}

Suppose $x_0$ and $x_1$ represent the initial concentrations for the molecules $X_0$ and $X_1$, respectively. From (\ref{eq:integral}) we have
\begin{eqnarray}
\label{eq:integral2}
&\ln [X_0]-\ln x_0=\ln [X_1] -\ln x_1 \nonumber \\
&\Rightarrow \frac{[X_0]}{x_0}=\frac{[X_1]}{x_1}.
\end{eqnarray}

Similarly, from the last two reactions in (\ref{divoed2}) we obtain
\begin{eqnarray}
\label{eq:integral3}
\frac{[Y_0]}{y_0}=\frac{[Y_1]}{y_1}
\end{eqnarray}
where $y_0$ and $y_1$ are the initial concentrations for the molecules $Y_0$ and $Y_1$, respectively.
The initial values for molecules $Z_0$ and $Z_1$ are zero and we can write
\begin{eqnarray}
\label{eq:Cratio}
\frac{[Z_1]}{[Z_0]+[Z_1]}=\frac{\frac{d[Z_1]}{dt}}{\frac{d[Z_0]}{dt}+\frac{d[Z_1]}{dt}}.
\end{eqnarray}

From the last two equations of (\ref{divoed}) we write
\begin{eqnarray}
\label{eq:Cratio2}
\frac{[Z_1]}{[Z_0]+[Z_1]}= \frac{[X_1][Y_0]+[X_1][Y_1]}{[X_0][Y_1]+2[X_1][Y_1]+[X_1][Y_0]}.
\end{eqnarray}
By substituting (\ref{eq:integral2}) and (\ref{eq:integral3}) into (\ref{eq:Cratio2}) we obtain
\begin{eqnarray}
\label{eq:Cratio3}
z&=&\frac{[Z_1]}{[Z_0]+[Z_1]} \nonumber \\
&=& \frac{\frac{y_0}{y_1}[X_1][Y_1]+[X_1][Y_1]}{\frac{x_0}{x_1}[X_1][Y_1]+2[X_1][Y_1]+\frac{y_0}{y_1}[X_1][Y_1]} \nonumber \\
&=&\frac{\frac{y_0}{y_1}+1}{\frac{x_0}{x_1}+2+\frac{y_0}{y_1}}\nonumber \\
&=&\frac{x_1(y_0+y_1)}{x_1(y_0+y_1)+y_1(x_0+x_1)}\nonumber \\
&=&\frac{\frac{x_1}{x_0+x_1}}{\frac{x_1}{x_1+x_0}+\frac{y_1}{y_1+y_0}}=\frac{x}{x+y}.
\end{eqnarray}

\subsection{Implementations of Sigmoid Functions and Tangent Hyperbolic in Bipolar Format}
Implementation of sigmoid functions requires computing sigmoid($ax$) where $a$ can be greater than $1$. This computation is reformulated using a division operation. The divider of the previous subsection is used in this subsection.

Consider the implementation of sigmoid($2ax$) ($a>0$) with bioplar input and unipolar output as follows \cite{liu2016computing}:

\begin{eqnarray}
sigmoid(2ax)&=&\frac{1}{1+e^{-2ax}}\\
&=&\frac{1}{1+e^{-2a(2P_x-1)}}\label{equ4}\\
&=&\frac{1}{1+e^{-4aP_x} \cdot e^{2a}}\\
&=&\frac{e^{-2a}}{e^{-2a}+e^{-4aP_x}}.
\end{eqnarray}

\begin{figure}[htbp]
\vspace{-0.5em}
\centering
\resizebox{0.48\textwidth}{!}{
\includegraphics{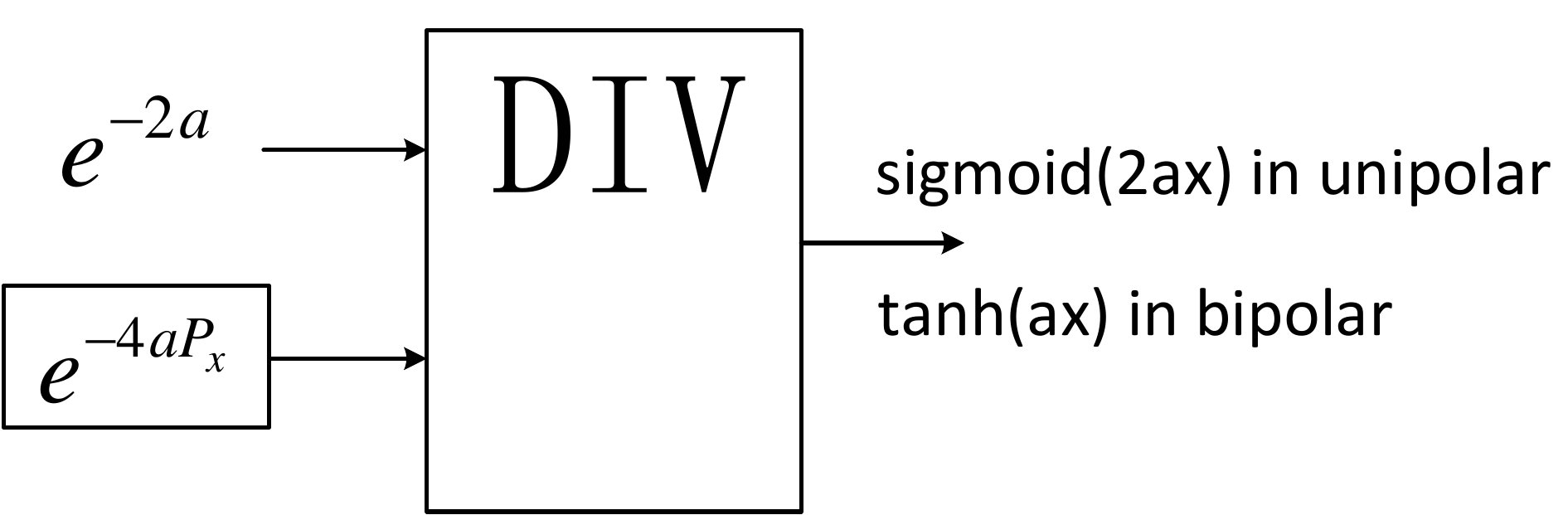}}
\caption{Stochastic implementation of $sigmoid(2ax)$ and $tanh(ax)$ with positive $a$.}
\label{sig2ax}
\vspace{-1em}
\end{figure}

In equation~(\ref{equ4}), $x$ is replaced by $2P_x-1$ where $P_x$ represents the unipolar value of the input bit stream while the output is also in unipolar format. So $sigmoid(2ax)$ can be implemented in unipolar logic. The molecular implementation is shown in Fig.~\ref{sig2ax}. In this design, $e^{-2a}$ $(a>0)$ is in the range of [0,1] which can be represented in unipolar format by using a pair of molecular types. The $e^{-4aP_x}$ is implemented using the proposed method in Section \Romannum{3}.A based on molecular reactions. Then $sigmoid(2ax)$ can be implemented by using the division function presented in Section \Romannum{3}.B with two inputs $e^{-2a}$ and $e^{-4aP_x}$.

Consider $tanh(ax)$ with positive $a$ described as follows:

\begin{eqnarray}
tanh(ax)&=&\frac{1-e^{-2ax}}{1+e^{-2ax}}\nonumber\\
&=&2\cdot \frac{1}{1+e^{-2ax}}-1\nonumber\\
&=&2\cdot sigmoid(2ax)-1\nonumber.
\end{eqnarray}

Given $-1\leq x\leq 1$, $sigmoid(2ax)$ is in the range $[0,1]$ represented in unipolar format. So if $sigmoid(2ax)=\frac{[S_1]}{[S_0]+[S_1]}$, then $tanh(ax)=2\cdot \frac{[S_1]}{[S_0]+[S_1]}-1=\frac{[S_1]-[S_0]}{[S_0]+[S_1]}$. Recall the definition of bipolar format $x=\frac{[X_1]-[X_0]}{[X_0]+[X_1]}$, where $[X_0]$ and $[X_1]$ represent the corresponding concentrations of the assigned molecular types while $x$ represents the bipolar value. We can observe that both functions can be implemented using the same molecular circuit shown in Fig.~\ref{sig2ax}, where the output concentrations in unipolar representation compute $sigmoid(2ax)$ whereas the output in bipolar representation computes $tanh(ax)$. Molecular sigmoids and molecular tangent hyperbolic functions have not been presented before. These are typically used as activation functions of perceptrons and ANNs.

\begin{figure}[htbp]
\vspace{-0em}
\centering
\resizebox{0.48\textwidth}{!}{
\includegraphics{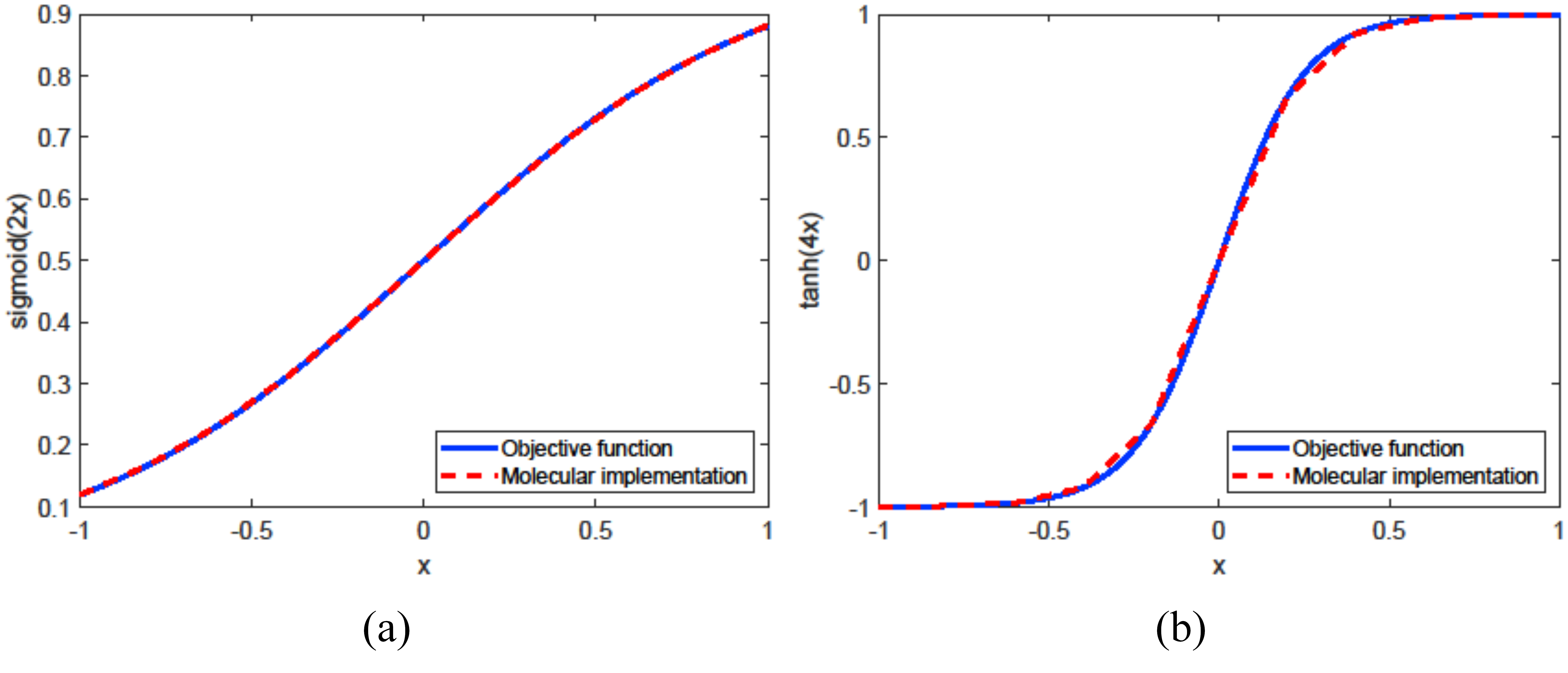}}
\caption{Simulation results of bipolar (a) $sigmoid(2x)$ and (b) $tanh(4x)$ implemented by using molecular reactions shown in Figs.~\ref{base} and \ref{div} with the proposed design shown in Fig.~\ref{sig2ax}.}
\label{sigres}
\vspace{-1em}
\end{figure}

Fig.~\ref{sigres} shows the simulation results of $sigmoid(2x)$ and $tanh(4x)$ using the proposed approach shown in Fig.~\ref{sig2ax} with \nth{5}-order truncated Maclaurin expansion of $e^{-x}$. We show CRNs of $sigmoid(2x)$ in Supplementary Section S.3.

\subsection{Computing Scaled Inner Products}
Notice that inner product ($\sum_{i=1}^{N}w_ix_i$) cannot be directly implemented since the result might not be bounded by $-1$ and $1$, which violates the constraint of bipolar format. Therefore, scaled versions of inner product are needed. In this section, two molecular approaches to computing scaled inner product of two inputs vectors in bipolar format are presented. 

\subsubsection{Inner Products Scaled by the Number of Inputs}
\begin{figure}[htbp]
\vspace{-0.8em}
\centering
\resizebox{0.3\textwidth}{!}{%
\includegraphics{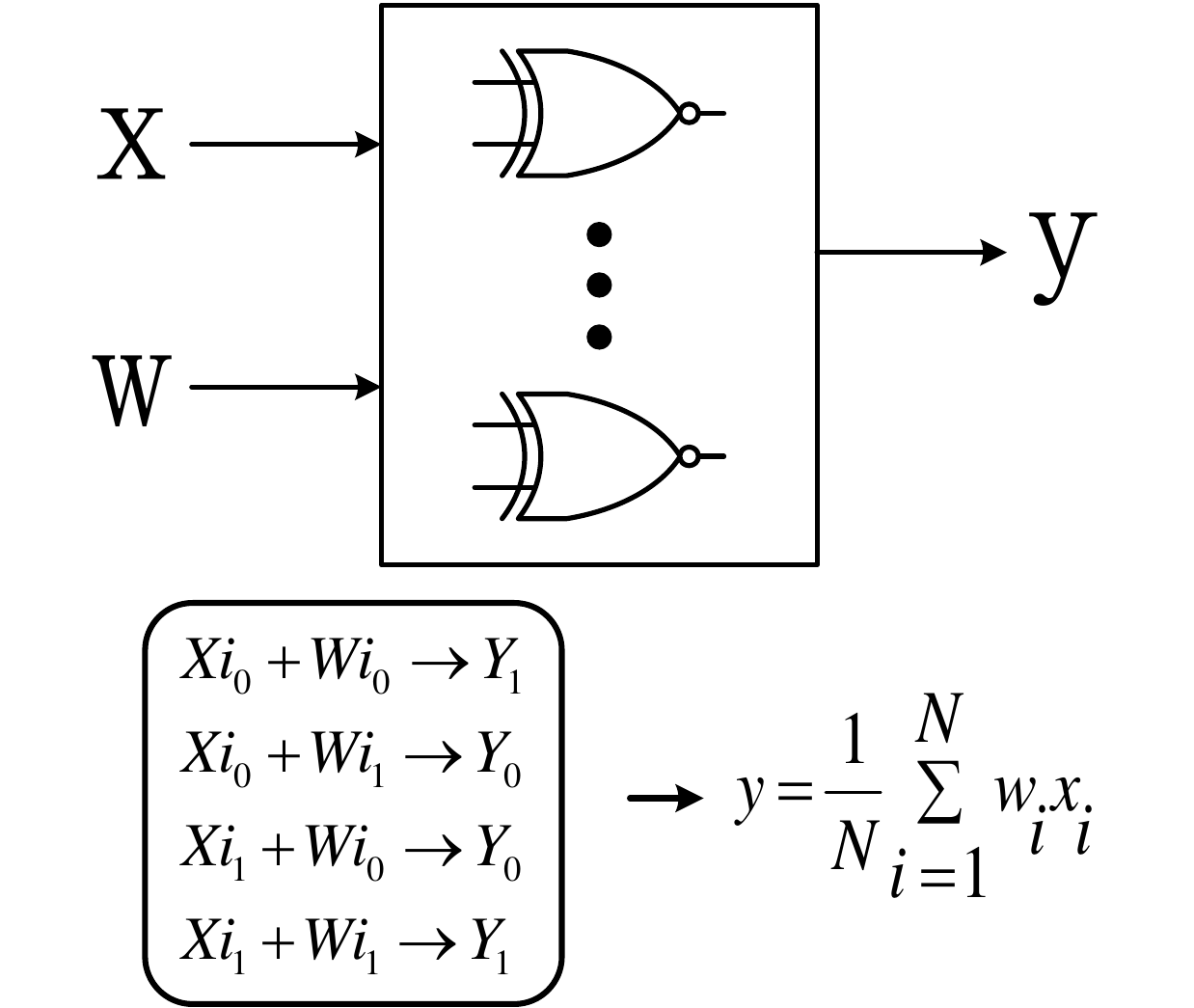}}
\caption{The inner product unit \Romannum{1}. This unit calculates $y=\frac{1}{N}\sum_{i=1}^{N}w_ix_i$, the scaled inner product of two input vectors $x$ and $w$ where each element is in bipolar fractional representation.}
\label{inner1}
\vspace{-1em}
\end{figure}

Given $x=\begin{bmatrix}x_1\\x_2\\:\\x_N\end{bmatrix}$ and $w=\begin{bmatrix}w_1\\w_2\\:\\w_N\end{bmatrix}$ as the two input vectors where each element is in bipolar format, we can compute the inner product functions scaled by the number of inputs $N$ with $4N$ molecular reactions as shown in Fig.~\ref{inner1} \cite{salehi2018computing}. Each of the four reactions is related to a bipolar \texttt{Mult} unit shown in Fig.~\ref{base}(c) with two corresponding inputs, $x_i$ and $w_i$. Fig.~\ref{inner1} also lists the proposed molecular reactions, where $i=1, 2, \cdots N$. Notice that $-1\leq x_i\leq 1$, $-1\leq w_i\leq 1$ must be guaranteed for feasible design. So if $x_i=\frac{[Xi_1]-[Xi_0]}{[Xi_1]+[Xi_0]}$ and $w_i=\frac{[Wi_1]-[Wi_0]}{[Wi_1]+[Wi_0]}$ then $y=\frac{[Y_1]-[Y_0]}{[Y_1]+[Y_0]}=\frac{1}{N}\sum_{i=1}^{N}w_ix_i$. A proof of the functionality of the molecular inner product in Fig.~\ref{inner1} is described in Section S.5 of the Supplementary Information in \cite{salehi2018computing}.

\subsubsection{Inner Products Scaled by the Sum of Absolute Value of Weights}
\begin{figure}[htbp]
\vspace{-0.8em}
\centering
\resizebox{0.48\textwidth}{!}{%
\includegraphics{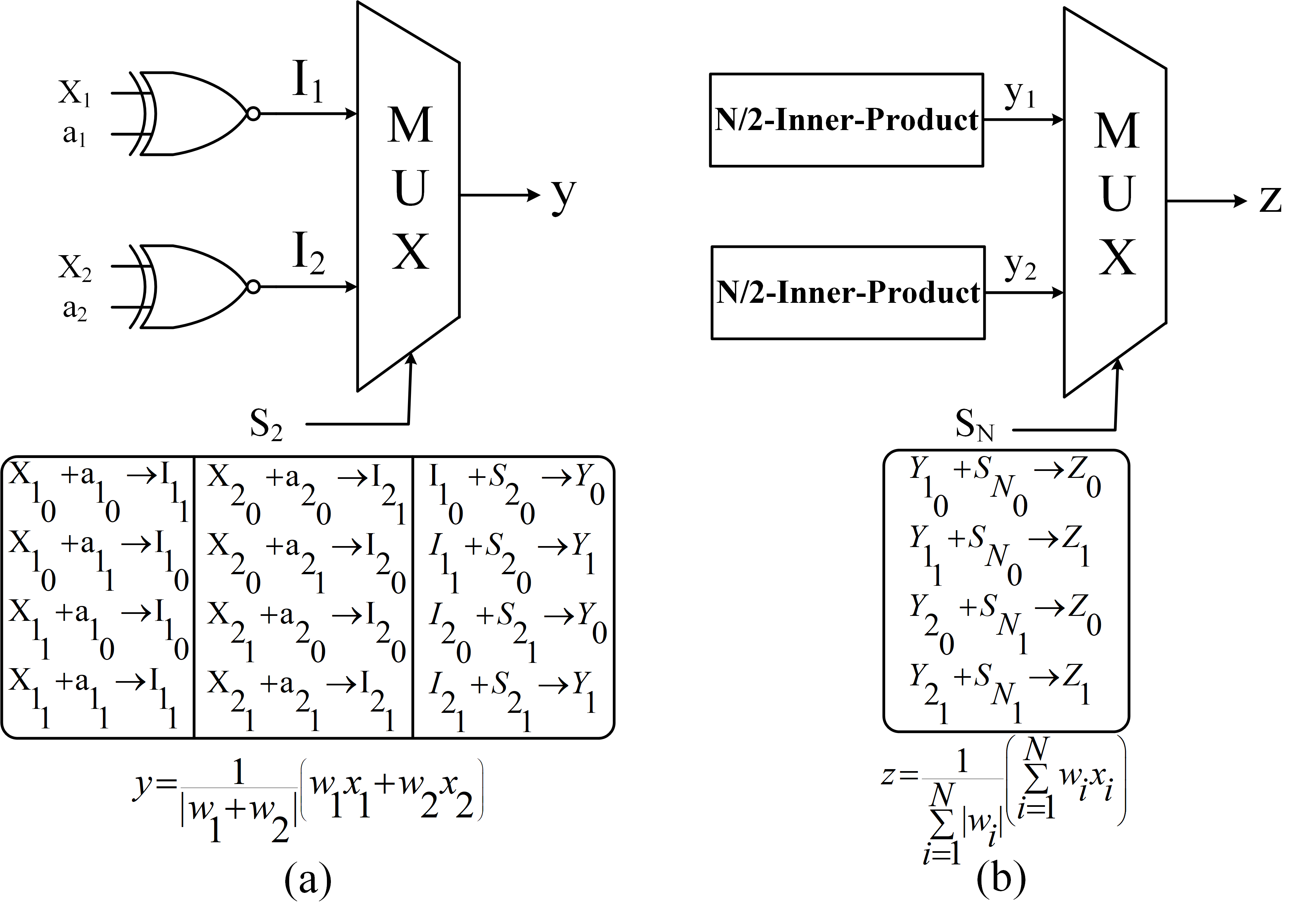}}
\caption{The inner product unit \Romannum{2} (a) with $2$-dimensional input vectors that calculates $y=\frac{1}{|w_1|+|w_2|}(w_1x_1+w_2x_2)$ and (b) with $N$-dimensional input vectors that calculates $z=\frac{1}{\sum_{i=1}^{N}|w_i|}(\sum_{i=1}^{N}w_ix_i)$ based on two $\frac{N}{2}$-dimensional inner products. In both cases, each element $x_i$ is in bipolar fractional representation and each element $w_i$ takes arbitrary values..}
\label{inner2base}
\vspace{-1em}
\end{figure}

In most ANNs, trained weights need not be guaranteed to lie between $-1$ and $1$ when the result of the inner product function scaled by the number of inputs as shown in Fig.~\ref{inner1} might be out of range for bipolar representation. The stochastic implementation of inner product scaled by sum of the absolute weights was proposed in \cite{chang2013architectures,liu2015lattice}. This method has no limitation on the range of elements in one input vector while the elements in the other input vector must be in range $[-1,1]$. Consider the basic condition when the dimension of the input vectors is $2$, where $x=\begin{bmatrix}x_1\\x_2\end{bmatrix}$ $(-1 \leq x_i\leq 1)$ and $w=\begin{bmatrix}w_1\\w_2\end{bmatrix}$ $(w_i\in \textbf{R})$. In this approach, the \texttt{MUX} units are used to perform addition and the \texttt{Mult} units are used to perform multiplication as shown in Fig.~\ref{inner2base}. Select signal of multiplexer is given by: $s_2=\frac{|w_1|}{|w_1|+|w_2|}$.

Fig.~\ref{inner2base}(a) describes a two-input inner product where $x_1$ and $x_2$ are in bipolar format and the pre-calculated coefficient $s_2$ is in unipolar format. Fig.~\ref{inner2base}(a) also lists the proposed molecular reactions. Besides, $a_1$ and $a_2$ represent signs of $w_1$ and $w_2$, respectively. If $w_i>0$, then $a_i=\frac{[{a_i}_1]-[{a_i}_0]}{[{a_i}_1]+[{a_i}_0]}=1$. Otherwise, $a_i=\frac{[{a_i}_1]-[{a_i}_0]}{[{a_i}_1]+[{a_i}_0]}=-1$. So given $X_1=\frac{[{X_1}_1]-[{X_1}_0]}{[{X_1}_1]+[{X_1}_0]}$, $X_2=\frac{[{X_2}_1]-[{X_2}_0]}{[{X_2}_1]+[{X_2}_0]}$ and $S_2=\frac{[{S_2}_1]}{[{S_2}_1]+[{S_2}_0]}$, then final output is given by: $y=\frac{[Y_1]-[Y_0]}{[Y_1]+[Y_0]}=\frac{1}{|w_1|+|w_2|}(w_1x_1+w_2x_2)$. Note that a $4$-input inner product can be computed recursively using two $2$-input inner products at first-level and then another $2$-input inner product at second-level. The recursive formulation leads to a tree-based design \cite{liu2015lattice}.

An inner product with $N$-dimensional input vectors can be implemented based on two inner products with $\frac{N}{2}$-dimensional input vectors as shown in Fig.~\ref{inner2base}(b). Consider the stochastic implementation of $z=\frac{1}{\sum_{i=1}^{N}|w_i|}\sum_{i=1}^{N}w_ix_i$, which can be written as follows:

\begin{eqnarray}
y_1&=&\frac{[{Y_1}_1]-[{Y_1}_0]}{[{Y_1}_1]+[{Y_1}_0]}=\frac{1}{\sum_{i=1}^{\frac{N}{2}}|w_i|}\sum_{i=1}^{\frac{N}{2}}w_ix_i\nonumber\\
y_2&=&\frac{[{Y_2}_1]-[{Y_2}_0]}{[{Y_2}_1]+[{Y_2}_0]}=\frac{1}{\sum_{i=\frac{N}{2}+1}^{N}|w_i|}\sum_{i=\frac{N}{2}+1}^{N}w_ix_i\nonumber\\
s_{N}&=&\frac{[{S_N}_1]}{[{S_N}_1]+[{S_N}_0]}=\frac{\sum_{i=1}^{\frac{N}{2}}|w_i|}{\sum_{i=1}^{N}|w_i|}\nonumber\\
z&=&\frac{[Z_1]-[Z_0]}{[Z_1]+[Z_0]}=s_{N}\cdot y_1+(1-s_{N})\cdot y_2.
\label{innerequ}
\end{eqnarray}

Notice that equation~(\ref{innerequ}) computes the scaled inner product ($z$) of two $N$-dimensional input vectors, $x$ and $w$, by using a \texttt{MUX} unit with the inputs $y_1$ and $y_2$ of two inner-product functions with corresponding halves of the input vectors while the select signal of the multiplexer is given by: $s_{N}=\frac{\sum_{i=1}^{\frac{N}{2}}|w_i|}{\sum_{i=1}^{N}|w_i|}$. Fig.~\ref{inner2base} shows the corresponding circuit diagram and molecular reactions for the implementation of $\frac{1}{\sum_{i=1}^{N}|w_i|}\sum_{i=1}^{N}w_ix_i$ in bipolar format using equation~(\ref{innerequ}). Inner product functions with odd-dimensional input vectors can also be implemented using the approach shown in Fig.~\ref{inner2base}(b) by adding an extra input feature with zero weight.

\section{Mapping Molecular Computing System to DNA}
Abstract chemical reaction networks (CRNs) described by molecular reactions can be mapped to DNA strand displacement (DSD) reactions as shown in \cite{salehi2018computing}. The DSD reactions based on toehold mediation  was primarily introduced by Yurke \textit{et al.} in~\cite{yurke2000dna}. A framework that can implement arbitrary molecular reactions with no more than two reactants by linear, double-stranded DNA complexes was proposed by Chen in~\cite{chen2013programmable}. Notice that our computational units are all built based on molecular reactions with at most two reactants. We simulate the ANN classifier for the EEG signal classification by using the software package provided by Winfree's team at Caltech \cite{soloveichik2010dna}. More details of mapping bimolecular reactions to DSD are described in Section S.8 of the Supplementary Information of \cite{salehi2018computing}.

\section{Molecular and DNA Perceptrons with Binary Inputs and Non-Binary Weights}
\begin{figure}[htbp]
\vspace{-0em}
\centering
\resizebox{0.48\textwidth}{!}{
\includegraphics{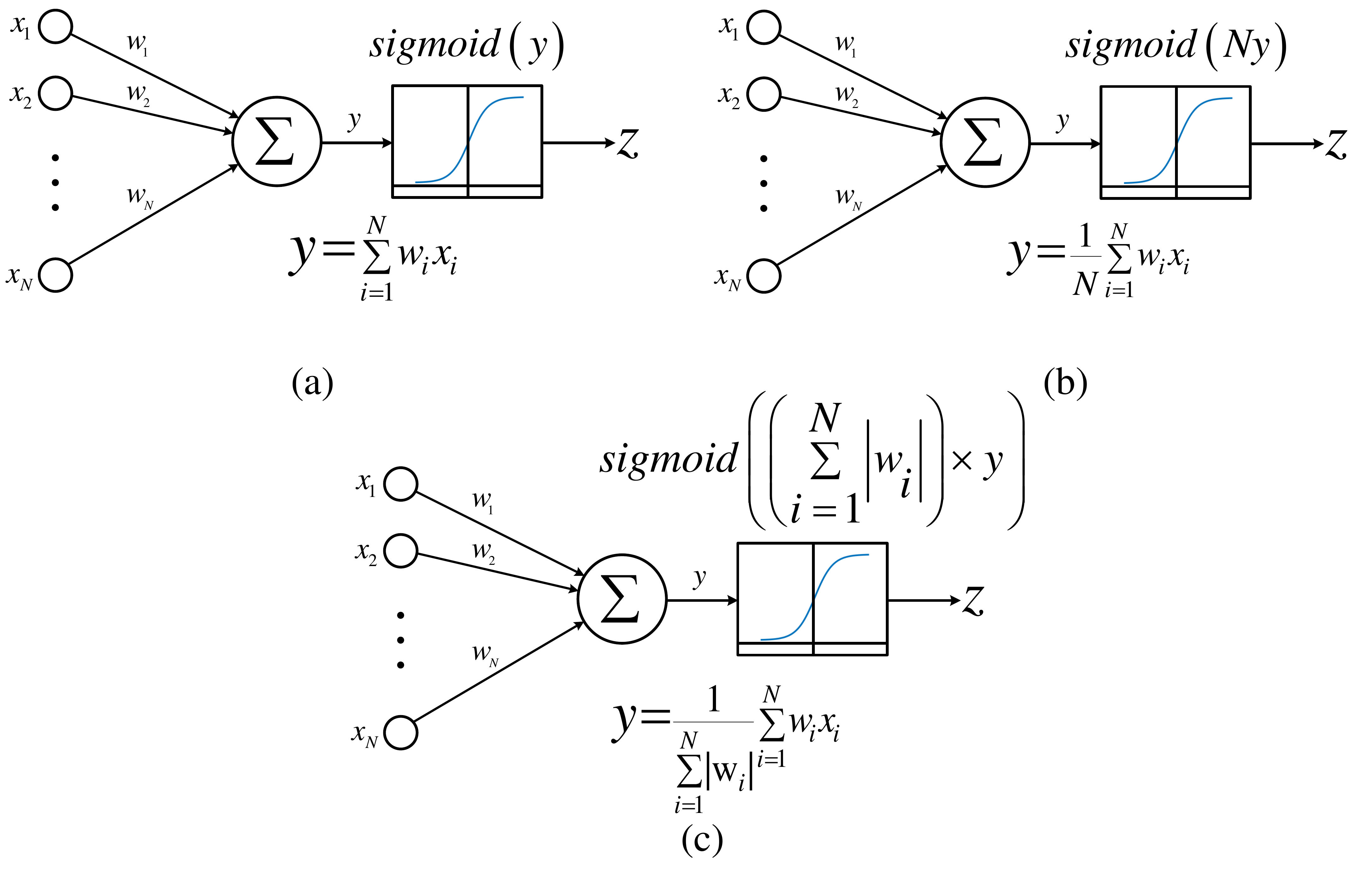}}
\caption{Molecular perceptron. (a) A standard perceptron that computes $sigmoid(\sum_{i=1}^{N}w_ix_i)$, (b) the molecular perceptron that computes $sigmoid(N \cdot (\frac{1}{N}\sum_{i=1}^{N}w_ix_i))$ and (c) the molecular perceptron that computes $sigmoid(\sum_{i=1}^{N}|w_i| \cdot (\frac{1}{\sum_{i=1}^{N}|w_i|}\sum_{i=1}^{N}w_ix_i))$.}
\label{per}
\vspace{-1em}
\end{figure}

In this section, we present the implementation of a single-layer neural network, also called a perceptron, using molecular reactions. The perceptron is illustrated in Fig.~\ref{per}(a) where the weighted sum $y=\sum_{i=1}^{N}w_ix_i$ is computed first. The final output $z$ is then computed as $z=sigmoid(y)$. The molecular inner products cannot compute the inner product exactly, but can compute only a scaled version of the inner product. Figs.~\ref{per}(b) and (c) illustrate two approaches to implement a molecular perceptron. In Fig.~\ref{per}(a), the inner product is scaled down by the number of inputs, $N$. In Fig.~\ref{per}(c), the inner product is scaled down by the sum of the absolute values of the weights $\sum_{i=1}^{N}|w_i|$. The molecular sigmoid, therefore, must compute the sigmoid of the scaled-up version of the input using the molecular $sigmoid(2ax)$ presented in Section \Romannum{3}. The molecular perceptrons shown in Figs.~\ref{per}(b) and (c) perform the same computation as the standard perceptron shown in Fig.~\ref{per}(a). The implementation of computing sigmoid of the weighted sum {\em scaled} by the number of inputs was presented in \cite{salehi2018computing}; such a sigmoid cannot be used for regression applications. The proposed molecular perceptrons in Figs.~\ref{per}(b) and (c) compute the exact sigmoid and can be used for regression applications.

\begin{figure}[htbp]
\vspace{-1em}
\centering
\resizebox{0.48\textwidth}{!}{
\includegraphics{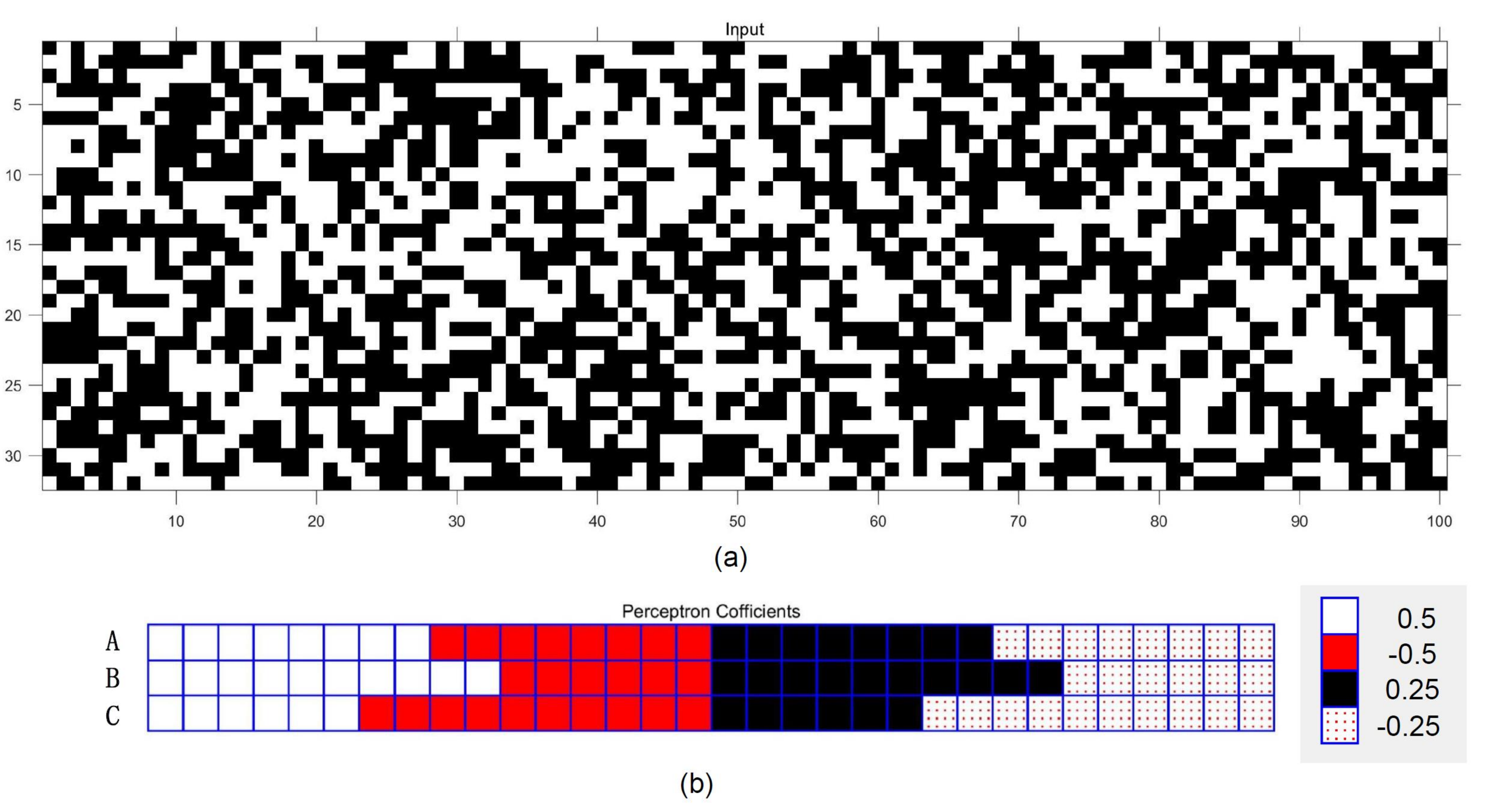}}
\caption{Inputs, weights of three perceptrons, denoted A, B, C. (a) Inputs to the perceptron: each column represents an input vector containing 32 binary inputs in this $32\times 100$ block matrix. Each black block corresponds to a $0$ and each white block represents a $1$. (b) Weights: the weights for the three perceptrons are illustrated. Each row represents the $32$ weights for one perceptron. These weights are divided into $4$ parts and correspond to $1/2$, $-1/2$, $1/4$, $-1/4$ from left to right. The figure is taken from \cite{salehi2018computing}.}
\label{pertwo}
\vspace{-1em}
\end{figure}

Three perceptrons are simulated with $N=32$ using the $32$ coefficients as described in \cite{salehi2018computing}. Notice that there are $100$ sets of inputs and each set consists of $32$ randomly selected binary numbers as shown in Fig.~\ref{pertwo}(a). However, the inputs of the proposed molecular perceptron don't have to be binary, but should be no less than $-1$ and no greater than $1$. Fig.~\ref{pertwo}(b) shows the weights for the 3 perceptrons, denoted A, B and C. Each weight, $1/2$, $-1/2$, $1/4$ and $-1/4$, occurs $8$ times in Perceptron A. The same weights occur $10$, $6$, $10$ and $6$ times, respectively, in Perceptron B; and $6$, $10$, $6$ and $10$ times, respectively, in Perceptron C. 

% Each weighted input either activates or inhibits the \textit{state} of the perceptron. Then Perceptron A has equal number of inputs that either activate or inhibit the state whereas Perceptron B has more inputs that activate the state, and Perceptron C has more inputs that inhibit the state. Since they have equally likely binary inputs, the expectations of the weighted sum for the Perceptrons A, B, and C are $0$, $1.5$ and $-1.5$, respectively. Then the expected sigmoid values for the three perceptrons are $0.5$, $0.8175$, and $0.1825$, respectively.

Three perceptrons are simulated by using the two methods shown in Figs.~\ref{per}(b) and (c). The first method cascades the inner product units scaled by the number of inputs and $sigmoid(32x)$ which can be implemented by setting $a=16$ in the approach illustrated in Fig.~\ref{sig2ax}. The second method cascades the inner product units scaled by the sum of absolute weights, $12$; $sigmoid(12x)$ can be implemented by setting $a=6$ in the approach illustrated in Fig.~\ref{sig2ax}. Note that an approach to compute $sigmoid(x)$ with implicit format conversion was presented in~\cite{li2017neural}; however, this approach is not applicable here as we need to compute $sigmoid(ax)$ with $a$ greater than $1$ and not $sigmoid(x)$. Various activation functions are used in neural networks, we show how Rectified Linear Unit and softmax function can be synthesized using molecular computing via fractional coding in Section \Romannum{7}.

\begin{figure}[htbp]
\vspace{-0em}
\centering
\resizebox{0.48\textwidth}{!}{
\includegraphics{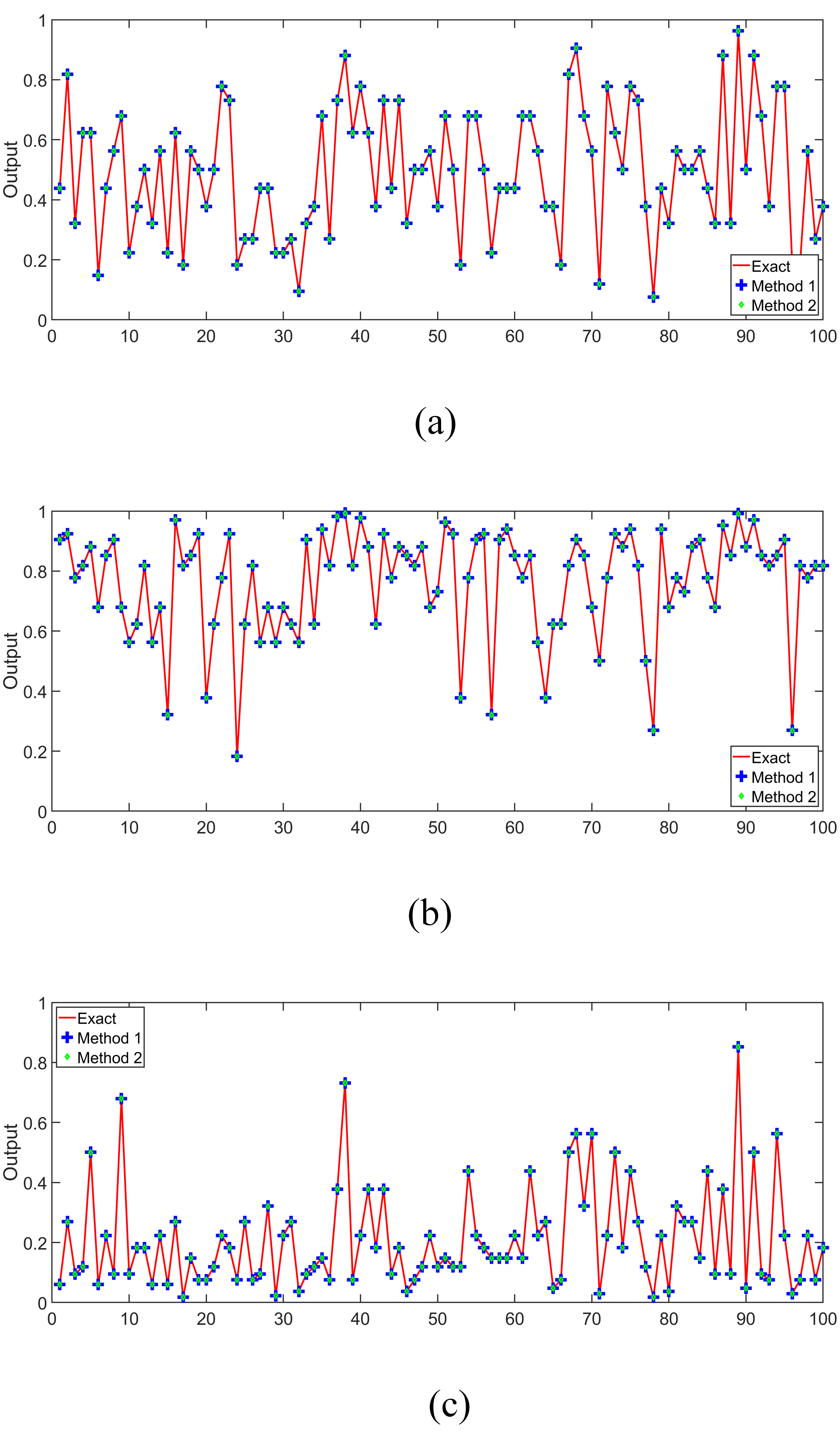}}
\caption{Exact and molecular outputs of three perceptrons: (a) Perceptron A, (b) Perceptron B, (c) Perceptron C. The $x$ axis represents $100$ random input vectors.}
\label{perress}
\vspace{-1em}
\end{figure}

\begin{figure}[htbp]
\vspace{-0em}
\centering
\resizebox{0.48\textwidth}{!}{
\includegraphics{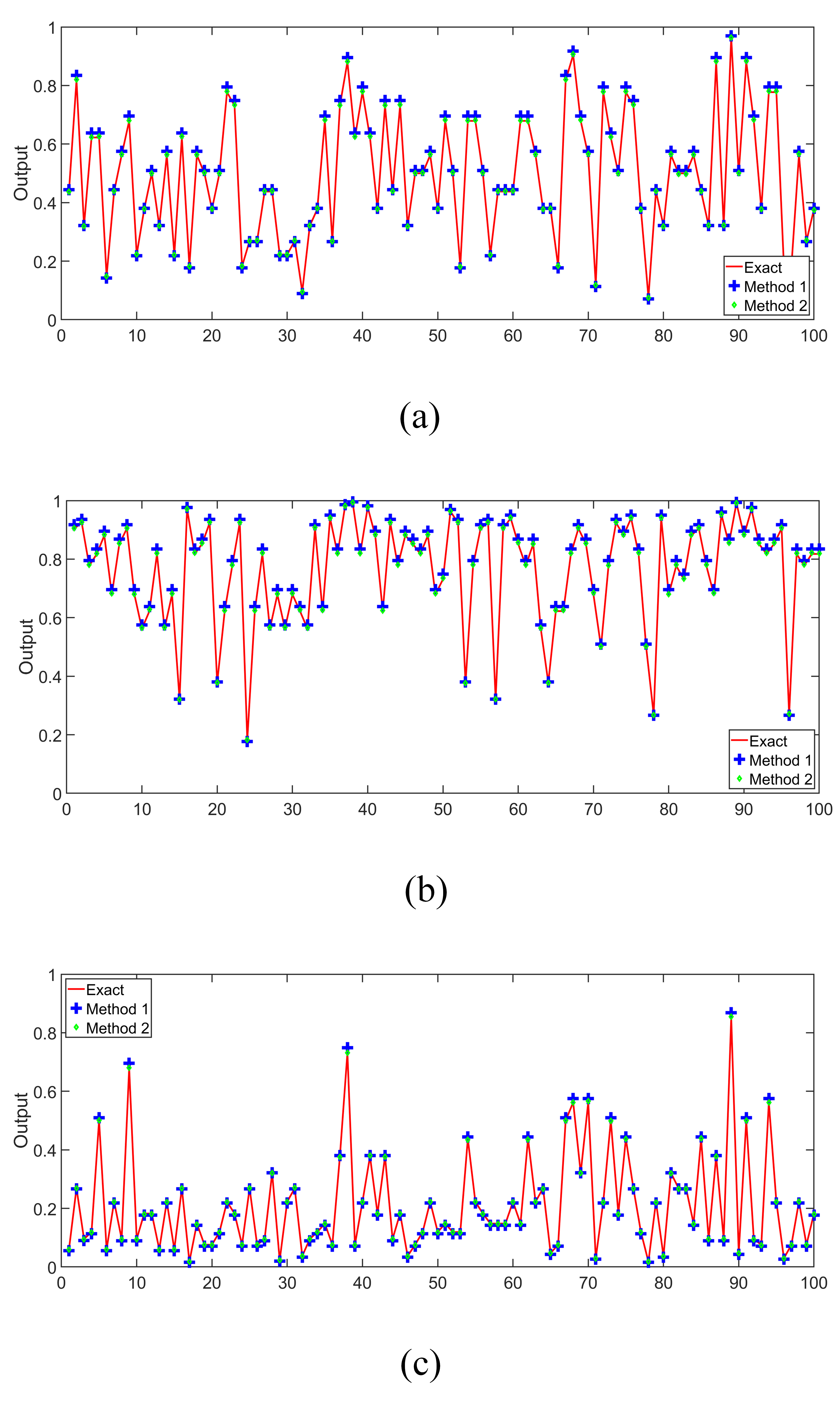}}
\caption{Exact and molecular outputs of three perceptrons: (a) Perceptron A, (b) Perceptron B, (c) Perceptron C. The $x$ axis represents $100$ random input vectors.}
\label{perressdna}
\vspace{-1em}
\end{figure}

The molecular simulation results of the three perceptrons are shown in Figs.~\ref{perress}(a)-(c), respectively. The bimolecular reactions are mapped to DNA as described in \cite{soloveichik2010dna}. The DNA simulation results of the three perceptrons are shown in Figs.~\ref{perressdna}(a)-(c), respectively. The red line illustrates the target outputs of $100$ input vectors; the blue crosses show the outputs of the perceptron as shown in Fig.~\ref{per}(b) where the scaling of the inner product is the number of the inputs (referred to as Method-1) and the green diamonds show the outputs of the perceptron as shown in Fig.~\ref{per}(c) where the scaling of the inner product is the sum of the absolute value of the coefficients (referred to as Method-2). The horizontal axis in Fig.~\ref{perress} represents the index of the input vector and the vertical axis shows the exact and molecular sigmoid values. The mean square error, \textit{MSE}, is defined as:

\begin{gather*}
\centering
MSE = \frac{1}{100} \sum_{j=1}^{100} \left | z(j) - \widehat{z}(j)\right |^2
\end{gather*}
where $z(j) = sigmoid(\sum_{i=1}^{N}w_ix_{i}[j])$ represents the exact value, $\widehat{z}(j)$ represents the simulation result, $x_{i}[j]$ represents the $i^{\text{th}}$ bit position of input vector $j$, and $w_i$ is the $i^{\text{th}}$ weight. The mean square error values for the molecular and DNA simulations are listed in Table~\ref{my-label}. The proposed Method-2 achieves less error than the Method-1.

\begin{table*}[]
\caption{Mean sqaure errors for the three perceptrons using molecular reactions and DNA.}
\centering
\resizebox{0.8\textwidth}{!}{
\begin{tabular}{|c|c|c|c|c|}
\hline
           & \multicolumn{2}{c|}{Molecular} & \multicolumn{2}{c|}{DNA} \\ \hline
Perceptron & Method 1       & Method 2      & Method 1    & Method 2   \\ \hline
A          & $1.78486\times 10^{-7}$ & $2.67643\times 10^{-8}$ & $1.22267\times 10^{-4}$ & $4.05697\times 10^{-6}$ \\ \hline
B          & $1.3643\times 10^{-7}$  & $2.56013\times 10^{-8}$ & $1.96767\times 10^{-4}$ & $7.57364\times 10^{-6}$ \\ \hline
C          & $7.29132\times 10^{-8}$ & $7.66375\times 10^{-9}$ & $3.69588\times 10^{-5}$ & $1.89104\times 10^{-6}$ \\ \hline
\end{tabular}
}
\label{my-label}
\end{table*}

\section{Molecular and DNA Implementation of ANN Classifiers}
Classification using a simple perceptron may be possible using a $sigmoid(x)$ function with scaled inputs. Thus, the prior work~\cite{salehi2018computing} may suffice to act as a classifier for a perceptron. However, such a sigmoid function as in~\cite{salehi2018computing} cannot be used as an activation function for neurons in the  hidden layer of an ANN. These neurons need to compute exact sigmoid values. Our paper describes a molecular ANN with one hidden layer for a seizure prediction application. The activation functions of the hidden layer neurons do require computing $sigmoid(x)$ without scaling of inputs, and thus act as regressors. The outputs of these regressors are fed to the output neuron. Thus, the contribution of this paper is significant in the context of molecular ANNs.
\subsection{Architecture of the Molecular Implementation}

\begin{figure}[htbp]
\vspace{-0em}
\centering
\resizebox{0.37\textwidth}{!}{
\includegraphics{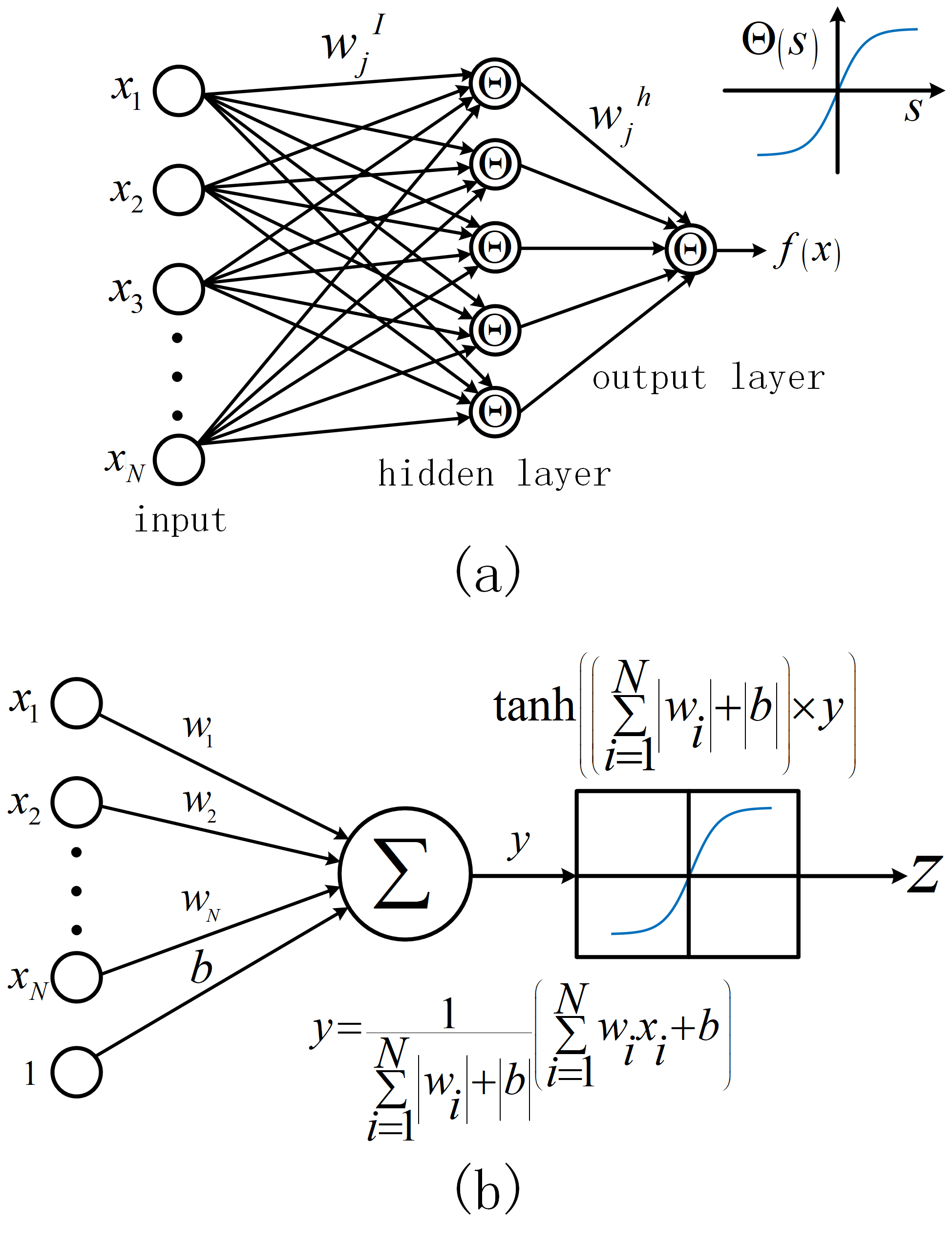}}
\caption{(a) An artificial neural network (ANN) model with one hidden layer, (b) computation kernel in a neuron implemented in molecular reactions.}
\label{ann}
\vspace{-1em}
\end{figure}

Consider the ANN with one hidden layer as shown in Fig.~\ref{ann}(a). Each neuron computes a weighted sum of the neuron values from the previous layer followed by an activation function ($\Theta$). Thus, each neuron contains an inner-product and a tangent hyperbolic function as shown in Fig.~\ref{ann}(b). The inner-product $\frac{1}{\sum_{i=1}^{N}|w_i|+|b|}\sum_{i=1}^{N}(w_ix_i+b)$ is computed at node $y$ by using the approach of inner product unit \Romannum{2} as shown in Fig.~\ref{inner2base} with $N+1$ inputs.

Notice that the molecular implementation of tangent hyperbolic functions $tanh(ax)$ is illustrated in Fig. \ref{sig2ax}. The result computed at node $y$ is a scaled version of the original weighted sum. However the output of the neuron implemented by molecular reactions is the same as the output of a conventional implementation while $tanh(ax)$ is implemented using $a=round(\sum_{i=1}^{N}|w_i|+|b|)$ instead of the original $tanh(x)$. 

\subsection{EEG Signal Classification using Molecular ANN Classifier}
\begin{figure}[htbp]
\vspace{0em}
\centering
\resizebox{0.4\textwidth}{!}{%
\includegraphics{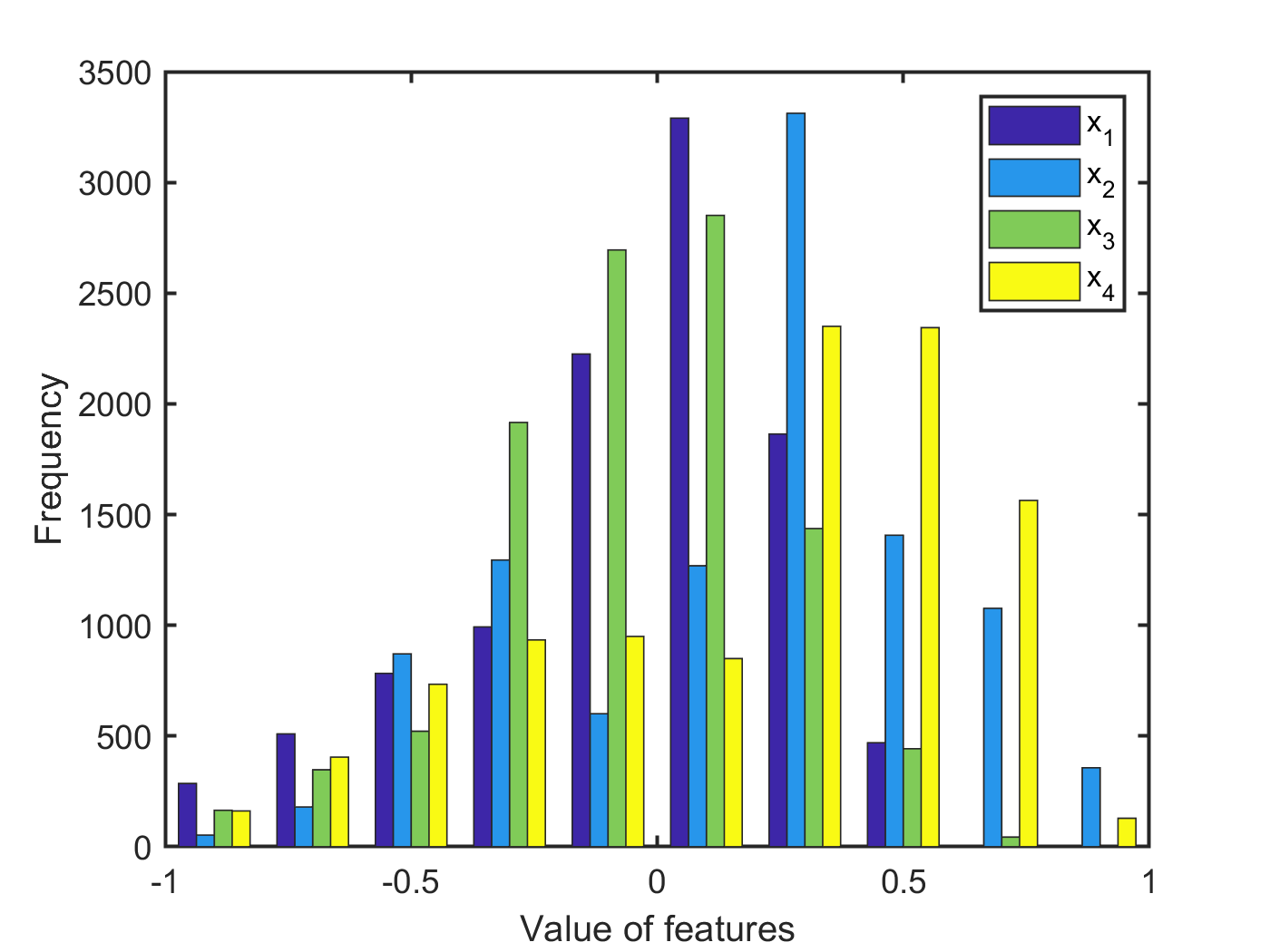}}
\caption{The histogram of input features after linear mapping.}
\label{his}
\vspace{-1em}
\end{figure}

\begin{figure*}[htbp]
\vspace{-0em}
\centering
\resizebox{0.75\textwidth}{!}{
\includegraphics{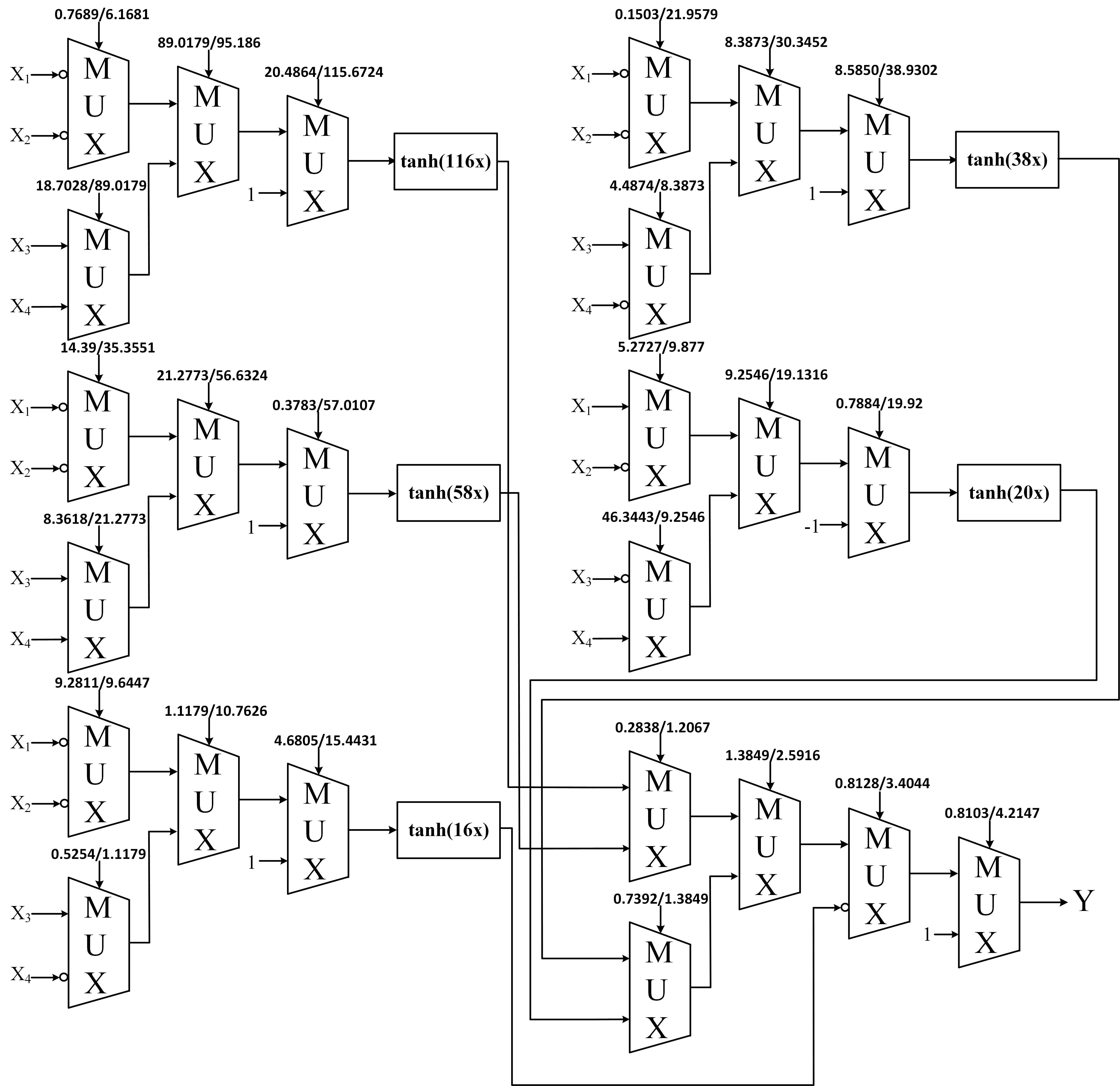}}
\caption{Molecular implementation of the ANN model.}
\label{annall}
\vspace{-1em}
\end{figure*}

The molecular ANN classifier is tested using an ANN with one hidden layer containing five neurons and four neurons for the input layer trained for an application for seizure prediction from electroencephalogram (EEG) signals \cite{zhang2016low}. The data from one human patient from the UPenn/Mayo Seizure Prediction Contest sponsored by the American Epilepsy Society and hosted by Kaggle is used for training the ANN \cite{kaggle}. In prior work, the same data was used to design an ANN using stochastic logic \cite{liu2016machine}. The testing data contains 10422 samples and each sample contains a vector with $4$ features ($x={x_1, x_2, x_3, x_4}$) and a bias term, $b$. Since the range of the input features should be $[-1, 1]$ under the constraint of bipolar format representation, a linear mapping is performed on the input features. Consider the input data as a $10422\times 4$ matrix $X$, where the number of rows ($10422$) and columns ($4$) represent the number of inputs data samples and the number of features, respectively. The linear mapping is performed for all samples as follows:

\begin{equation}
\centering
X_{ij}\Leftarrow \frac{2(X_{ij}-min(X_i))}{max(X_i)-min(X_i)}, \text{for $1\leq i\leq 4$, $1\leq j\leq 10422$} \nonumber
\end{equation}
where $max(X_i)$ and $min(X_i)$ represent the maximum and minimum magnitudes of the $i^{\text{th}}$ feature among all $10422$ samples. After this linear mapping, each element in a column of the input matrix $X$ has mean $0$ and the dynamic range of $[-1 ,1]$. The histogram of $41688$ features from 10422 $4$-dimensional feature vectors after linear mapping is shown in Fig.~\ref{his}. The threshold for the finial classification is zero. Table~\ref{wab} lists the weight matrices and bias vectors of the optimized ANN model, where $w^I$ represents the connection weight matrix of the input-hidden layer connection, $w^h$ represents the hidden layer-output connection, $b_h$ represents the bias column vector for the hidden neurons, and $b_o$ is the bias for the output neuron. Fig.~\ref{annall} shows the molecular implementation of the ANN model. This circuit includes $58$ \texttt{Mult} units, $25$ \texttt{NMult} units, $25$ \texttt{MUX} units and $5$ division units. Table~\ref{number} states the number of reactants and the number of reactions for the molecular ANN. The length of molecular simulation time is $50$ hours. Using the coefficients of the ANN and actual data from a patient with $10422$ samples, the confusion matrices for the ideal classification results from MATLAB using the trained ANN and the molecular classification results are shown in Tables~\ref{Matlabres} and \ref{Mores}, respectively. Here the ground truth is the actual label of the data from the patient. The ideal MATLAB results in Table~\ref{Matlabres} are the best achievable by an ANN with one hidden layer with five neurons. To achieve better accuracy, an ANN with more neurons in the hidden layer or more number of hidden layers should be trained. 

\begin{table}[H]
\centering
\caption{Number of Reactants and Reactions for Molecular and ANN DNA Implementations.}
\begin{tabular}{|c|c|c|}
\hline
          & Molecular ANN & DNA ANN \\ \hline
Reactants &  456 &  5184   \\ \hline
Reactions &  678 &  2260   \\ \hline
\end{tabular}
\label{number}
\end{table}

\begin{table*}[]
\caption{Weights and bias of the optimized ANN model.}
\centering
\resizebox{0.7\textwidth}{!}{
\begin{tabular}{|c|c|c|c|c|c|c|}
\hline
\multicolumn{5}{|c|}{Input - Hidden layer connections} & \multicolumn{2}{c|}{Hidden layer - Output connections} \\ \hline
\multicolumn{4}{|c|}{Weights}               & Bias     & Weights           & Bias                               \\ \hline
$w_{j1}^I$        & $w_{j2}^I$        & $w_{j3}^I$       & $w_{j4}^I$       & $b_{hj}$      & $w_j^h$                 & $b_o$                                 \\ \hline
-5.3992   & -0.7689   & 70.3151  & 18.7028  & 20.4864  & 0.9229            & \multirow{5}{*}{-0.8103}           \\ \cline{1-6}
-20.9651  & -14.3900  & 12.9155  & 8.3618   & 0.3783   & 0.2838            &                                    \\ \cline{1-6}
-21.8076  & -0.1503   & 3.8999   & -4.4874  & 8.5850   & 0.6457            &                                    \\ \cline{1-6}
4.6043    & -5.2727   & -2.9103  & 6.3443   & -0.7884  & 0.7392            &                                    \\ \cline{1-6}
-0.3636   & -9.2811   & 0.5925   & -0.5254  & 4.6805   & -0.8128           &                                    \\ \hline
\end{tabular}
}
\label{wab}
\end{table*}

\begin{table}[H]
\caption{The confusion matrix of classification using MATLAB.}
\begin{tabular}{ccccc}
\multicolumn{1}{l}{}                                                                         & \multicolumn{1}{l}{}                   & \multicolumn{2}{c}{Predicted}                                                   & \multicolumn{1}{l}{}            \\ \cline{3-4}
                                                                                             & \multicolumn{1}{c|}{}                  & \multicolumn{1}{c|}{\textbf{Positive}} & \multicolumn{1}{c|}{\textbf{Negative}} &                                 \\ \cline{2-5} 
\multicolumn{1}{c|}{\multirow{2}{*}{\begin{tabular}[c]{@{}c@{}}Actual\\ Class\end{tabular}}} & \multicolumn{1}{c|}{\textbf{Positive}} & \multicolumn{1}{c|}{TP=5267}           & \multicolumn{1}{c|}{FN=124}            & \multicolumn{1}{c|}{TPR=0.9770} \\ \cline{2-5} 
\multicolumn{1}{c|}{}                                                                        & \multicolumn{1}{c|}{\textbf{Negative}} & \multicolumn{1}{c|}{FP=368}            & \multicolumn{1}{c|}{TN=4663}           & \multicolumn{1}{c|}{TNR=0.9269} \\ \cline{2-5} 
\multicolumn{1}{c|}{}                                                                        & \multicolumn{1}{c|}{ACC=0.9528}        & \multicolumn{1}{c|}{PPV=0.9347}        & \multicolumn{1}{c|}{NPV=0.9741}        &                                 \\ \cline{2-4}
\end{tabular}
\label{Matlabres}
\end{table}

\begin{table}[H]
\caption{The confusion matrix of classification using molecular ANN.}
\begin{tabular}{ccccc}
\multicolumn{1}{l}{}                                                                         & \multicolumn{1}{l}{}                   & \multicolumn{2}{c}{Predicted}                                                   & \multicolumn{1}{l}{}            \\ \cline{3-4}
                                                                                             & \multicolumn{1}{c|}{}                  & \multicolumn{1}{c|}{\textbf{Positive}} & \multicolumn{1}{c|}{\textbf{Negative}} &                                 \\ \cline{2-5} 
\multicolumn{1}{c|}{\multirow{2}{*}{\begin{tabular}[c]{@{}c@{}}Actual\\ Class\end{tabular}}} & \multicolumn{1}{c|}{\textbf{Positive}} & \multicolumn{1}{c|}{TP=5250}           & \multicolumn{1}{c|}{FN=141}            & \multicolumn{1}{c|}{TPR=0.9738} \\ \cline{2-5} 
\multicolumn{1}{c|}{}                                                                        & \multicolumn{1}{c|}{\textbf{Negative}} & \multicolumn{1}{c|}{FP=357}            & \multicolumn{1}{c|}{TN=4674}           & \multicolumn{1}{c|}{TNR=0.9290} \\ \cline{2-5} 
\multicolumn{1}{c|}{}                                                                        & \multicolumn{1}{c|}{ACC=0.9522}        & \multicolumn{1}{c|}{PPV=0.9363}        & \multicolumn{1}{c|}{NPV=0.9707}        &                                 \\ \cline{2-4}
\end{tabular}
\label{Mores}
\end{table}

Comparing the accuracy results in Table~\ref{Matlabres} with that in Table~\ref{Mores}, we can find that the accuracy (ACC) of the proposed molecular ANN is $0.9522$, which is $0.06\%$ less than the ACC of the ideal results. Table~\ref{MoresbM} shows the confusion matrix of molecular classification results using the ANN model where the classification results from MATLAB represent the ground truth. It can be observed that the performance of molecular ANN using linear mapping for input data is close to the ideal results from MATLAB.

\begin{table}[H]
\caption{The confusion matrix of classification using molecular ANN based on ideal classification results.}
\begin{tabular}{ccccc}
\multicolumn{1}{l}{}                                                                              & \multicolumn{1}{l}{}                   & \multicolumn{2}{c}{Predicted}                                                   & \multicolumn{1}{l}{}            \\ \cline{3-4}
                                                                                                  & \multicolumn{1}{c|}{}                  & \multicolumn{1}{c|}{\textbf{Positive}} & \multicolumn{1}{c|}{\textbf{Negative}} &                                 \\ \cline{2-5} 
\multicolumn{1}{c|}{\multirow{2}{*}{\begin{tabular}[c]{@{}c@{}}Matlab\\ Class\end{tabular}}} & \multicolumn{1}{c|}{\textbf{Positive}} & \multicolumn{1}{c|}{TP=5605}           & \multicolumn{1}{c|}{FN=30}             & \multicolumn{1}{c|}{TPR=0.9947} \\ \cline{2-5} 
\multicolumn{1}{c|}{}                                                                             & \multicolumn{1}{c|}{\textbf{Negative}} & \multicolumn{1}{c|}{FP=2}              & \multicolumn{1}{c|}{TN=4785}           & \multicolumn{1}{c|}{TNR=0.9996} \\ \cline{2-5} 
\multicolumn{1}{c|}{}                                                                             & \multicolumn{1}{c|}{ACC=0.9969}        & \multicolumn{1}{c|}{PPV=0.9996}        & \multicolumn{1}{c|}{NPV=0.993769}      &                                 \\ \cline{2-4}
\end{tabular}
\label{MoresbM}
\end{table}

\subsection{Architecture of the DNA Implementation}

\begin{table}[H]
\caption{The confusion matrix of classification using DSD reactions.}
\begin{tabular}{ccccc}
\multicolumn{1}{l}{}                                                                         & \multicolumn{1}{l}{}                   & \multicolumn{2}{c}{Predicted}                                                   & \multicolumn{1}{l}{}            \\ \cline{3-4}
                                                                                             & \multicolumn{1}{c|}{}                  & \multicolumn{1}{c|}{\textbf{Positive}} & \multicolumn{1}{c|}{\textbf{Negative}} &                                 \\ \cline{2-5} 
\multicolumn{1}{c|}{\multirow{2}{*}{\begin{tabular}[c]{@{}c@{}}Actual\\ Class\end{tabular}}} & \multicolumn{1}{c|}{\textbf{Positive}} & \multicolumn{1}{c|}{TP=5250}           & \multicolumn{1}{c|}{FN=141}            & \multicolumn{1}{c|}{TPR=0.9738} \\ \cline{2-5} 
\multicolumn{1}{c|}{}                                                                        & \multicolumn{1}{c|}{\textbf{Negative}} & \multicolumn{1}{c|}{FP=362}            & \multicolumn{1}{c|}{TN=4669}           & \multicolumn{1}{c|}{TNR=0.9280} \\ \cline{2-5} 
\multicolumn{1}{c|}{}                                                                        & \multicolumn{1}{c|}{ACC=0.9517}        & \multicolumn{1}{c|}{PPV=0.9355}        & \multicolumn{1}{c|}{NPV=0.9707}        &                                 \\ \cline{2-4}
\end{tabular}
\label{DNAres}
\end{table}

\begin{table}[H]
\caption{The confusion matrix of classification using DSD reactions based on molecular classification results.}
\begin{tabular}{ccccc}
\multicolumn{1}{l}{}                                                                         & \multicolumn{1}{l}{}                   & \multicolumn{2}{c}{Predicted}                                                   & \multicolumn{1}{l}{}            \\ \cline{3-4}
                                                                                             & \multicolumn{1}{c|}{}                  & \multicolumn{1}{c|}{\textbf{Positive}} & \multicolumn{1}{c|}{\textbf{Negative}} &                                 \\ \cline{2-5} 
\multicolumn{1}{c|}{\multirow{2}{*}{\begin{tabular}[c]{@{}c@{}}Actual\\ Class\end{tabular}}} & \multicolumn{1}{c|}{\textbf{Positive}} & \multicolumn{1}{c|}{TP=5590}           & \multicolumn{1}{c|}{FN=17}            & \multicolumn{1}{c|}{TPR=0.9970} \\ \cline{2-5} 
\multicolumn{1}{c|}{}                                                                        & \multicolumn{1}{c|}{\textbf{Negative}} & \multicolumn{1}{c|}{FP=22}            & \multicolumn{1}{c|}{TN=4793}           & \multicolumn{1}{c|}{TNR=0.9954} \\ \cline{2-5} 
\multicolumn{1}{c|}{}                                                                        & \multicolumn{1}{c|}{ACC=0.9963}        & \multicolumn{1}{c|}{PPV=0.9961}        & \multicolumn{1}{c|}{NPV=0.9965}        &                                 \\ \cline{2-4}
\end{tabular}
\label{DNAresb}
\end{table}

Confusion matrices as shown in Tables~\ref{DNAres} and~\ref{DNAresb} present the classification results from DNA classifiers based on actual classification results and the molecular classification results from Section \Romannum{4}, respectively. We can see that the ACC of the proposed ANN using DSD reactions is 0.9517, which is only 0.11\% less than the ACC of the molecular classification. Table~\ref{number} also states the number of reactants and the number of reactions for the DNA ANN.

\section{Molecular and DNA Implementations of ReLU and Softmax Functions}
This section describes molecular and DNA reactions for implementing ReLU and softmax functions for artificial neural networks. Molecular implementation of a Gaussian kernel has been described in~\cite{liu2019computing} in the context of a support vector machine classifier.
\subsection{DNA Implementation for Rectified Linear Unit}
\begin{figure}
	\centering
	\includegraphics[width=2.2in]{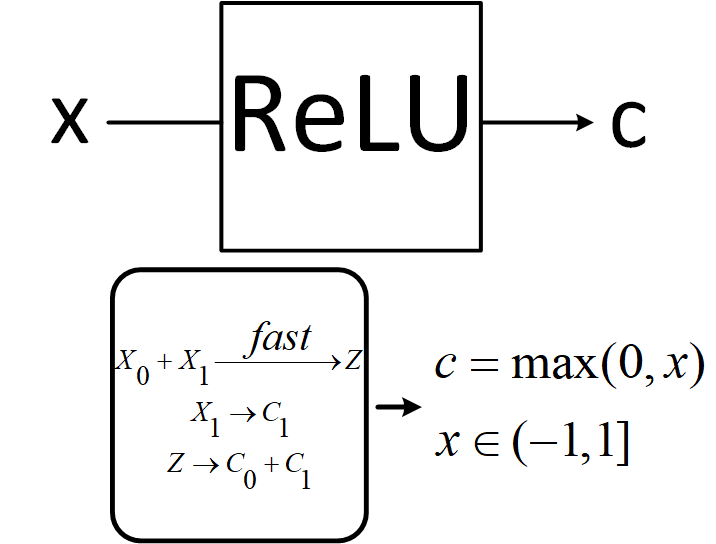}
\caption{Molecular rectified linear unit with input $x\in (-1,1]$. All the reactions have the same rate.}
\label{fig:relu}
\end{figure}

In the context of an artificial neural network, the ReLU activation function is defined as:
\begin{equation}
f(x)=x^+=max(0,x)
\end{equation}
where $x$ is the input to a neuron. Based on fractional coding, we propose a simple set of CRNs for implementing ReLU with inputs $x\in [-1,1]$. Notice that ReLU activation is not needed for unipolar inputs where the output is always the same as input. The molecular ReLU is described in Fig.~\ref{fig:relu}. We show some examples of ReLU in Supplementary Section S.2.1.

\begin{figure}
	\centering
	\includegraphics[width=2.3in]{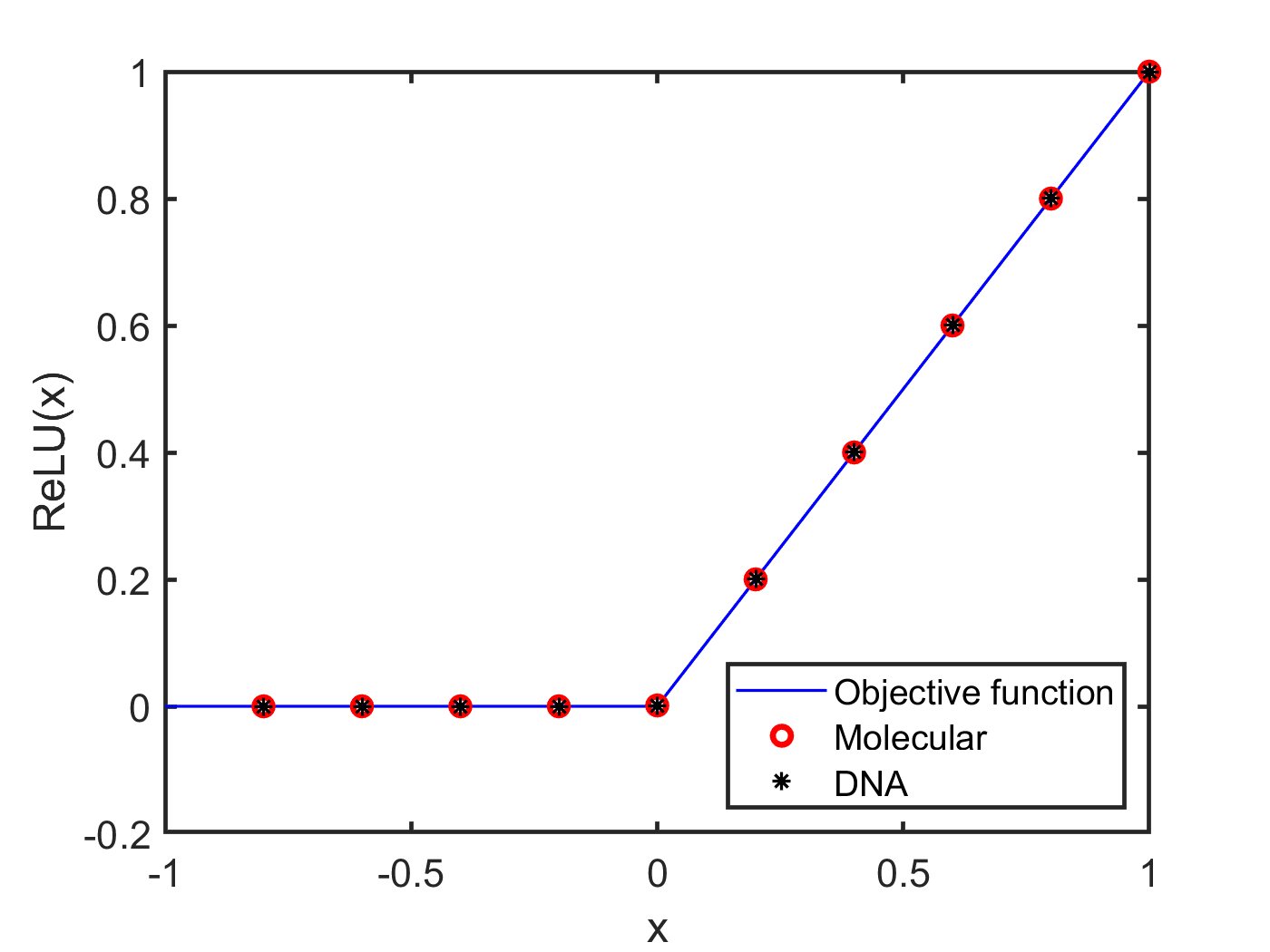}
\caption{Exact and computed values of the rectified linear unit. Blue lines: exact values, red circles: computed values using CRNs, black stars: computed values using DNA.}
\label{fig:relures}
\end{figure}

Fig.~\ref{fig:relures} shows the exact and simulated values of the rectified linear unit with bipolar inputs and bipolar outputs where the blue line represents the exact value, red circles represent the simulated values using CRNs, and black stars represent the simulated values using DNA.

\subsection{DNA Implementation for Softmax function}
Consider a standard softmax function with $3$ different classes with inputs $y_j$ for $j~=~1,2,3$. This function is defined by:
\begin{eqnarray}
\sigma(y_i)&=&\frac{e^{y_i}}{\sum_{j=1}^3 e^{y_j}}.\nonumber
\end{eqnarray}
The above can be reformulated by scaling the numerator and denominator by $e^{-1}$. The reformulates softmax is given by:
\begin{eqnarray}
\sigma(y_i)&=&\frac{e^{y_i}\cdot e^{-1}}{\sum_{j=1}^3 (e^{y_j}\cdot e^{-1})}\nonumber\\
&=&\frac{e^{y_i-1}}{\sum_{j=1}^3 e^{y_j-1}}\nonumber\\
&=&\frac{z_i}{z_1+z_2+z_3}.
\end{eqnarray}
where $z_j=e^{y_j-1}=\frac{[{Z_j}_1]}{[{Z_j}_1]+[{Z_j}_0]}$ for $j=1, 2, 3$. 

The stochastic implementation of $e^{(x-1)}$ with unipolar input and unipolar output has been described in Parhi~\cite{parhi2018stochastic} based on polynomial expansion described below.

\begin{eqnarray}
e^{(x-1)}&\approx&\frac{2}{e}(\frac{1}{2}+\frac{x}{2})+\frac{2}{3e}(\frac{3}{4}x^2+\frac{1}{4}x^3)\nonumber\\
&=&\frac{8}{3e}(\frac{3}{4}(\frac{1}{2}+\frac{x}{2})+\frac{1}{4}(\frac{3}{4}x^2+\frac{1}{4}x^3))\label{eqn:ex_1}.
\end{eqnarray}

Notice that the implementation of $e^{(x-1)}$ with bipolar input and unipolar output can be derived based on implicit format conversion. Let $x~=~2P_x - 1$ where $x$ is in bipolar and $P_x$ is in unipolar. Then we can write:

\begin{eqnarray}
e^{(x-1)}&=&e^{(2P_x-2)}\label{eqn:ex_1b1}\\
&=&(e^{(P_x)-1})^2\label{eqn:ex_1b2}.
\end{eqnarray}

\begin{figure}[htbp]
\vspace{-1em}
\centering
\resizebox{0.4\textwidth}{!}{
\includegraphics{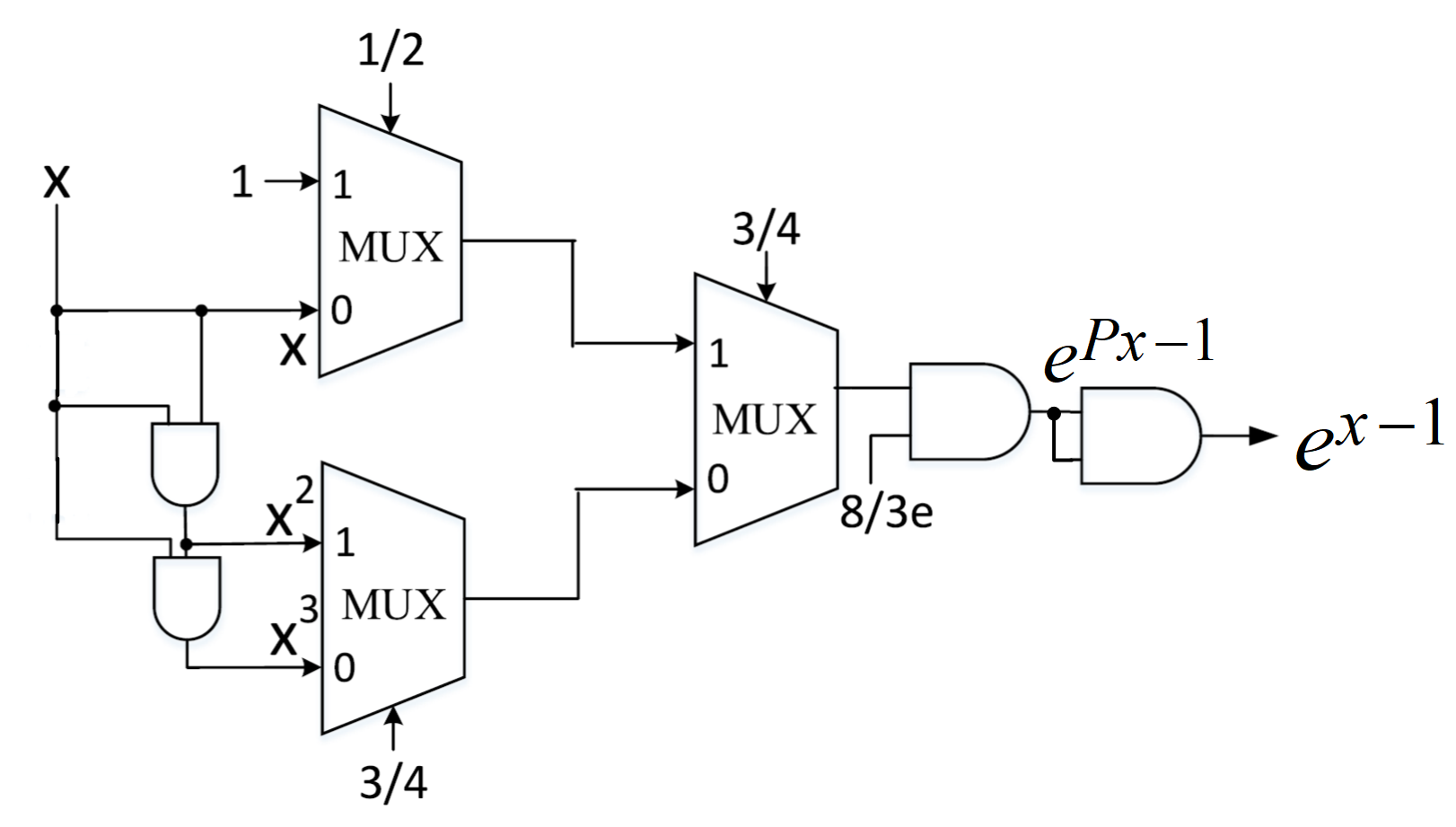}}
\caption{Stochastic implementation of $e^{(x-1)}$ with bipolar input.}
\label{ex1}
\vspace{-1em}
\end{figure}

In equation (\ref{eqn:ex_1b1}), $x$ is replaced by $2P_x-1$ where $P_x$ represents the unipolar value of the input bit stream while the output is also in unipolar logic. The square operation in equation (\ref{eqn:ex_1b2}) can be implemented by an \texttt{AND} unit with two identical inputs ($e^{(P_x)-1}$) that is computed by using equation (\ref{eqn:ex_1}). Fig.~\ref{ex1} shows the implementation of $e^{(x-1)}$ based on $e^{(Px-1)}$ and an \texttt{AND} gate.

Then the softmax function with $3$ inputs can be implemented by the following molecular reactions:

\begin{eqnarray}
\centering
{Z_1}_0+{Z_2}_0\rightarrow I_3 &\quad& {Z_1}_0+{Z_2}_1\rightarrow I_3 \nonumber\\
{Z_1}_1+{Z_2}_0\rightarrow I_3 &\quad& {Z_1}_1+{Z_2}_1\rightarrow I_3 \nonumber\\
{Z_1}_0+{Z_3}_0\rightarrow I_2 &\quad& {Z_1}_0+{Z_3}_1\rightarrow I_2 \nonumber\\
{Z_1}_1+{Z_3}_0\rightarrow I_2 &\quad& {Z_1}_1+{Z_3}_1\rightarrow I_2 \nonumber\\
{Z_2}_0+{Z_3}_0\rightarrow I_1 &\quad& {Z_2}_0+{Z_3}_1\rightarrow I_1 \nonumber\\
{Z_2}_1+{Z_3}_0\rightarrow I_1 &\quad& {Z_2}_1+{Z_3}_1\rightarrow I_1 \nonumber\\
{Z_1}_1+I_1 &\rightarrow& {Y_1}_1+{Y_2}_0+{Y_3}_0 \nonumber\\
{Z_1}_0+I_1 &\rightarrow& W \nonumber\\
{Z_2}_0+I_1 \rightarrow W &\quad& {Z_2}_1+I_1 \rightarrow W \nonumber\\
{Z_3}_0+I_1 \rightarrow W &\quad& {Z_3}_1+I_1 \rightarrow W \nonumber\\
{Z_2}_1+I_2 &\rightarrow& {Y_1}_0+{Y_2}_1+{Y_3}_0 \nonumber\\
{Z_2}_0+I_2 &\rightarrow& W \nonumber\\
{Z_1}_0+I_2 \rightarrow W &\quad& {Z_1}_1+I_2 \rightarrow W \nonumber\\
{Z_3}_0+I_2 \rightarrow W &\quad& {Z_3}_1+I_2 \rightarrow W \nonumber\\
{Z_3}_1+I_3 &\rightarrow& {Y_1}_0+{Y_2}_0+{Y_3}_1+{Z_3}_1+I_3. \nonumber \\
{Z_3}_0+I_3 &\rightarrow& W \nonumber\\
{Z_1}_0+I_3 \rightarrow W &\quad& {Z_1}_1+I_3 \rightarrow W \nonumber\\
{Z_2}_0+I_3 \rightarrow W &\quad& {Z_2}_1+I_3 \rightarrow W \nonumber
\end{eqnarray}

\begin{table}[]
\centering
\caption{Exact values and computed values of the softmax function with three inputs based on molecular and DNA.}
\begin{tabular}{|c|c|c|c|}
\hline
      & \multicolumn{3}{c|}{Softmax}    \\ \hline
input & Exact    & Molecular & DNA      \\ \hline
-0.2  & 0.195759 & 0.199504  & 0.199504 \\ \hline
0.3   & 0.322752 & 0.326513  & 0.326512 \\ \hline
0.7   & 0.481489 & 0.473983  & 0.473983  \\ \hline
\end{tabular}
\label{softres}
\end{table}

Mass-action kinetic equations for these reactions are discussed in Supplementary Information Section S.3.1. Table~\ref{softres} shows the exact values and computed values of the softmax function with three inputs ($-0.2,0.3,0.7$) based on the proposed molecular reactions and the corresponding DNA implementation. The molecular implementation of the softmax function with $K$ different classes is shown in Supplementary Information Section S.3.2.

\section{Conclusion}
This paper has shown that artificial neural networks can be synthesized using molecular reactions as well as DNA using fractional coding and bimolecular reactions. While an ANN with one hidden layer has been demonstrated, the approach applies to ANNs with multiple hidden layers. While theoretical feasibility has been demonstrated by simulations, any practical use of the proposed theory still remains to be validated. Despite lack of practical validation at this time, the theoretical advance is significant as it can quickly pave way for practical use either {\em in-vitro} or {\em in-vivo} when the technology becomes readily available. Future molecular sensing and computing systems can compute the features {\em in-situ} and these features can be used for classification using molecular artificial neural networks. Future work needs to be directed towards practical demonstration of the proposed theoretical framework. This paper has addressed molecular ANNs for inference applications. Inspired by~\cite{banda2013online}, a DNA perceptron that can {\em learn} was presented in~\cite{liu2019training}. We caution the reader that the high accuracy of the simulation of the chemical kinetics of the molecular systems may not reflect the accuracy of an experiment in a test tube for example.

\ifCLASSOPTIONcaptionsoff
  \newpage
\fi



%\bibliographystyle{IEEEtran}

\bibliographystyle{ieeetr} 
\bibliography{main}

\clearpage
\pagenumbering{arabic}
\setcounter{section}{0}

\subfile{bare2.tex}
\end{document}

%% file: bare2.tex
	\maketitle	
	
\section{Molecular Division unit}
\subsection{Division Input, Output Plots in DNA}
\begin{figure*}[!htb]
	\centering
	\includegraphics[height=10cm]{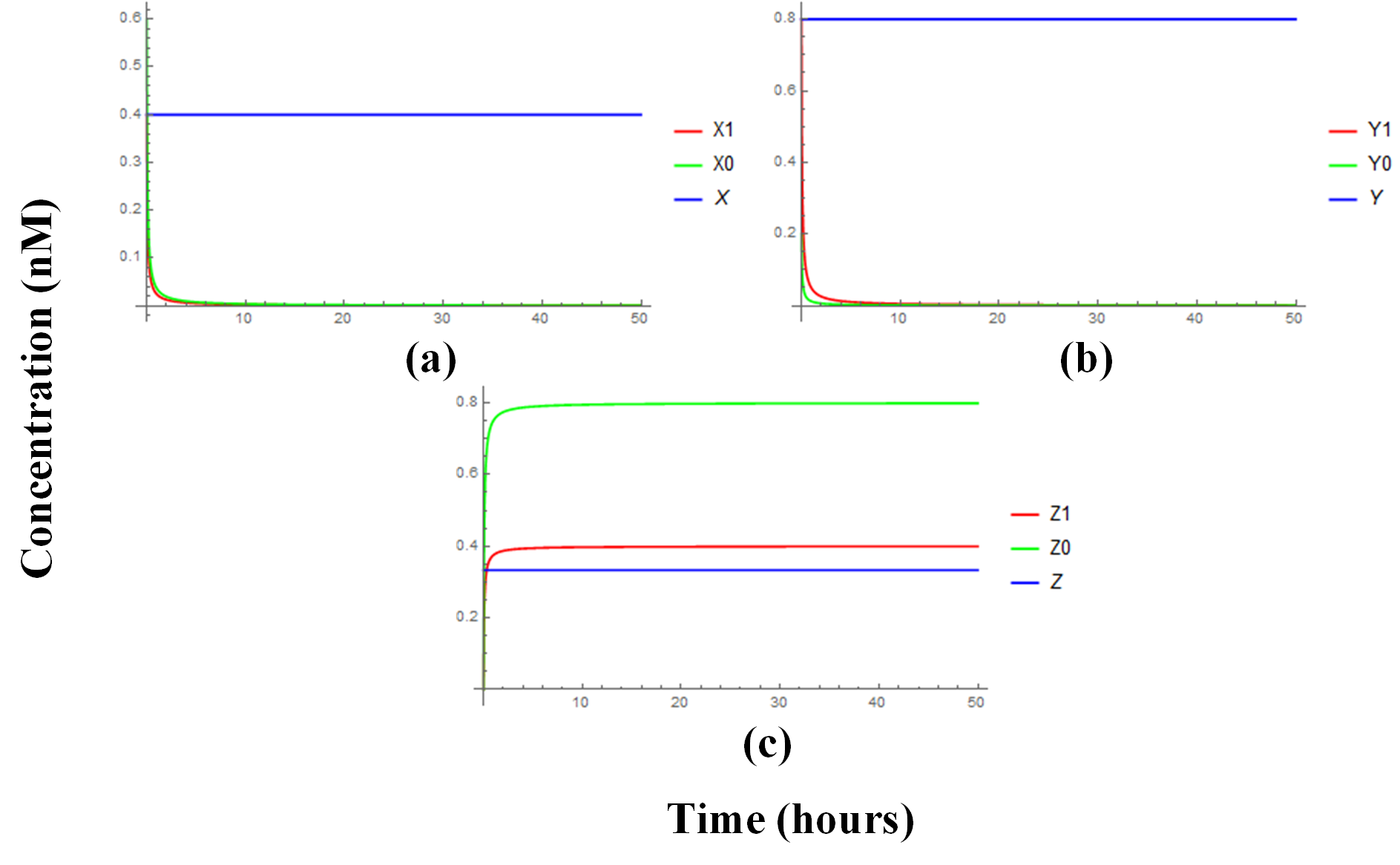}
	\vspace{-1em}
	\caption{\textbf{DNA simulation results.} The DNA reaction kinetics for a division unit with $x = 0.4$ and $y=0.8$: (a) Kinetics of input $x$ and its assigned molecules $X_1$ and $X_0$, (b) Kinetics of input $y$ and its assigned molecules $Y_1$ and $Y_0$, (c) Kinetics of output $z$ and its assigned molecules $Z_1$ and $Z_0$. }
	\label{fig:division}
\end{figure*}

In this section, the DNA reaction kinetics for the inputs and output of a division unit are shown in Fig.~\ref{fig:division}. As the proposed chemical reactions proceed, the concentrations of the input molecules ($X_0$, $X_1$, $Y_0$ and $Y_1$) decrease to zero while the output molecules ($Z_0$ and $Z_1$) reach their steady-state values. At equilibrium, the output $z=\frac{[Z_1]}{[Z_0]+[Z_1]}=\frac{0.4}{0.4+0.8}=\frac{1}{3}=\frac{x}{x+y}$ which is desired.

\subsection{Molecular and DNA MUX-DIV}

\begin{figure}[!htb]
	\centering
	\includegraphics[height=7cm]{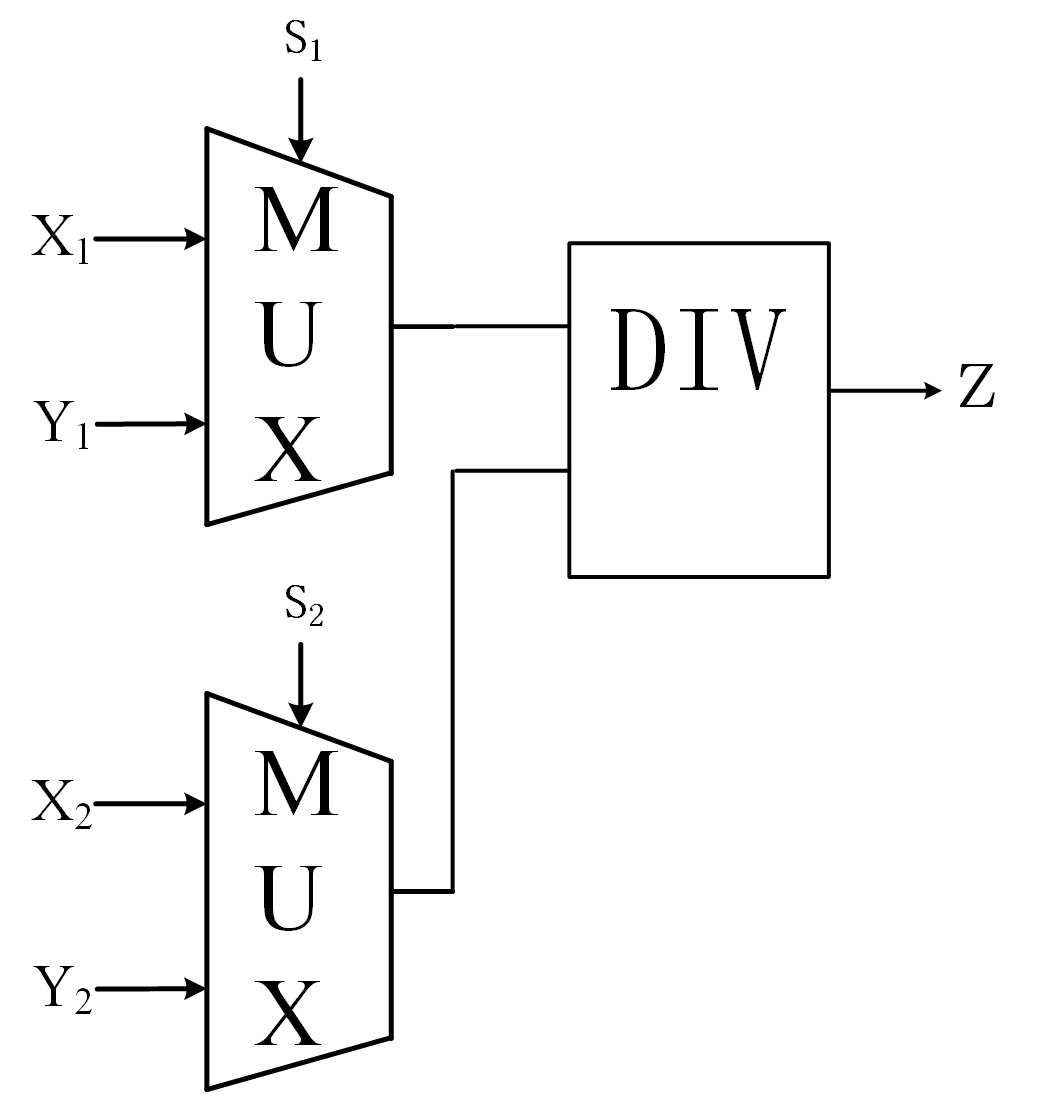}
	\vspace{-1em}
	\caption{Two \texttt{MUX} units cascaded by a division unit.}
	\label{fig:muxdiv}
\end{figure}

In this section, we prove the consistency between the CRN for MUX and the one for division unit by cascading two MUX units and a division unit as shown in Fig.~\ref{fig:muxdiv}. Then the molecular MUX-DIV can be implemented by the following molecular reactions:

\begin{eqnarray}
{X_1}_0+{S_1}_0&\rightarrow& {C_1}_0 \nonumber\\
{X_1}_1+{S_1}_0&\rightarrow& {C_1}_1 \nonumber\\
{Y_1}_0+{S_1}_1&\rightarrow& {C_1}_0 \nonumber\\
{Y_1}_1+{S_1}_1&\rightarrow& {C_1}_1 \nonumber\\
{X_2}_0+{S_2}_0&\rightarrow& {C_2}_0 \nonumber\\
{X_2}_1+{S_2}_0&\rightarrow& {C_2}_1 \nonumber\\
{Y_2}_0+{S_2}_1&\rightarrow& {C_2}_0 \nonumber\\
{Y_2}_1+{S_2}_1&\rightarrow& {C_2}_1 \nonumber\\
{C_1}_0+{C_2}_0&\rightarrow& W \nonumber\\
{C_1}_0+{C_2}_1&\rightarrow& {Z}_0 \nonumber\\
{C_1}_1+{C_2}_0&\rightarrow& {Z}_1 \nonumber\\
{C_1}_1+{C_2}_1&\rightarrow& {Z}_0+{Z}_1 \nonumber
\end{eqnarray}

Notice that four inputs ($X_1$, $Y_1$, $X_2$, $Y_2$), two select bits ($S_1$ and $S_2$) and the output ($Z$) are all in unipolar format. Table~\ref{table:muxdiv} shows the molecular and DNA results for MUX-DIV with one randomly selected input pair.

\begin{table*}[]
\centering
\caption{Molecular and DNA results of MUX-DIV with the proposed CRNs compared to their exact values.}
\resizebox{\textwidth}{!}{
\begin{tabular}{|c|c|c|c|c|c|c|c|c|c|c|c|c|c|c|c|c|}
\hline
\multicolumn{12}{|c|}{Initial Imput}                                                                                                                             & \multicolumn{5}{c|}{Final Output}                                         \\ \hline
\multicolumn{2}{|c|}{$X_1$}  & \multicolumn{2}{c|}{$Y_1$}  & \multicolumn{2}{c|}{$S_1$}  & \multicolumn{2}{c|}{$X_2$}  & \multicolumn{2}{c|}{$Y_2$}  & \multicolumn{2}{c|}{$S_2$}  & Exact    & \multicolumn{2}{c|}{Molecular} & \multicolumn{2}{c|}{DNA}      \\ \hline
\multicolumn{2}{|c|}{0.8} & \multicolumn{2}{c|}{0.1} & \multicolumn{2}{c|}{0.5} & \multicolumn{2}{c|}{0.3} & \multicolumn{2}{c|}{0.4} & \multicolumn{2}{c|}{0.7} & 0.548780 & \multicolumn{2}{c|}{0.548784}  & \multicolumn{2}{c|}{0.548784} \\ \hline
$[{X_1}_1]$          & $[{X_1}_0]$        & $[{Y_1}_1]$         & $[{Y_1}_0]$        & $[{S_1}_1]$         & $[{S_1}_0]$        & $[{X_2}_1]$         & $[{X_2}_0]$        & $[{Y_2}_1]$         & $[{Y_2}_0]$        & $[{S_2}_1]$         & $[{S_2}_0]$        &          & $[{Z_m}_1]$            & $[{Z_m}_0]$           & $[{Z_d}_1]$            & $[{Z_d}_0]$          \\ \cline{1-12} \cline{14-17} 
0.8          & 0.2        & 0.1         & 0.9        & 0.5         & 0.5        & 0.3         & 0.7        & 0.4         & 0.6        & 0.7         & 0.3        &          & 0.449699       & 0.369748      & 0.449699       & 0.369748     \\ \cline{1-12} \cline{14-17} 
\end{tabular}}
\label{table:muxdiv}
\end{table*}

\section{Molecular Rectified Linear Unit (ReLU)}
\subsection{Steady-State Analysis of ReLU}

Consider the three reactions for ReLU shown in Fig. 18 that compute the ReLU output, $c$, in bipolar representation. The input $x$ is also bipolar except that $x$ cannot be $-1$. When $x=-1$, $[C_0]$ and $[C_1]$ will always be $0$ where the output $c=\frac{[C_1]-[C_0]}{[C_0]+[C_1]}$ is undefined. Suppose $x_0$ and $x_1$ represent the initial concentrations for the molecules $X_0$ and $X_1$, respectively. If $x$ is negative ($[X_1]<[X_0]$), then $[X_0]=x_0-x_1, [X_1]=0, [Z]=x_1$ after the first reaction; the second reaction cannot occur due to the lack of $X_1$; the third reaction will convert all the $Z$ to $C_0$ and $C_1$, so $[C_0]=[C_1]=x_1$ at equilibrium. If $x$ is non-negative ($[X_1]\geq [X_0]$), then $[X_0]=0, [X_1]=x_1-x_0, [Z]=x_0$ after the first reaction; the second reaction will convert all the remaining $X_1$ to $C_1$; the third reaction will convert all the $Z$ to $C_0$ and $C_1$, so $[C_0]=x_0, [C_1]=x_1-x_0+x_0=x_1$ at equilibrium.

The output value, $c$, is given by:

\begin{align}
c&=\frac{[C_1]-[C_0]}{[C_1]+[C_0]}\nonumber \\
&=\left\{
\begin{array}{ccl}
\frac{x_1-x_1}{x_1+x_1}=0         &        & {-1 \leq x < 0}\\
\frac{x_1-x_0}{x_1+x_0}=x     &      & {\quad0 \leq x \leq 1}
\end{array} \right.\nonumber
\end{align}

\subsection{Molecular and DNA MUX-ReLU}

\begin{figure}[!htb]
	\centering
	\includegraphics[height=4cm]{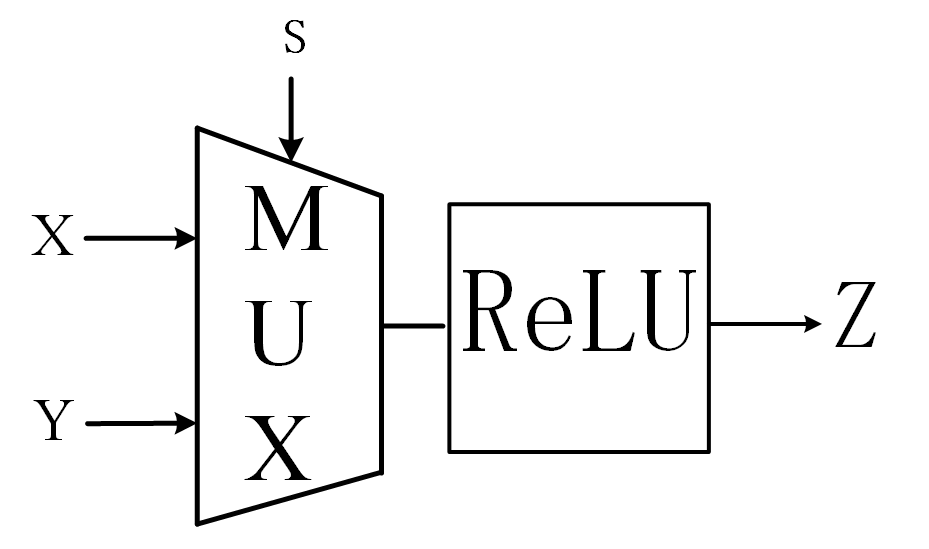}
	\vspace{-1em}
	\caption{A \texttt{MUX} unit cascaded by a ReLU unit. }
	\label{fig:muxrelu}
\end{figure}

In this section, we prove the consistency between the CRN for MUX and the one for ReLU by cascading a MUX unit and a ReLU unit as shown in Fig.~\ref{fig:muxrelu}. Then the molecular MUX-ReLU can be implemented by the following molecular reactions:

\begin{eqnarray}
X_0+S_0&\rightarrow& C_0 \nonumber\\
X_1+S_0&\rightarrow& C_1 \nonumber\\
Y_0+S_1&\rightarrow& C_0 \nonumber\\
Y_1+S_1&\rightarrow& C_1 \nonumber\\
C_0+C_1&\underrightarrow{fast}& F \nonumber\\
C_1&\rightarrow& Z_1 \nonumber\\
F&\rightarrow& Z_0+Z_1 \nonumber
\end{eqnarray}

Notice that the two inputs ($X$ and $Y$) and the output ($Z$) are all in bipolar format while the select bit ($S$) is in unipolar format.
Table~\ref{table:muxrelu} shows the molecular and DNA results for MUX-ReLU with two different input pairs.

\begin{table*}[]
\centering
\caption{Molecular and DNA results of MUX-ReLU with the proposed CRNs compared to their exact values.}
\resizebox{\textwidth}{!}{
\begin{tabular}{|c|c|c|c|c|c|c|c|c|c|c|c|}
\hline
\multirow{5}{*}{Exp. 1}                       & \multicolumn{6}{c|}{Initial Input}                                                                                                                              & \multicolumn{5}{c|}{Final Output}                                                                                                               \\ \cline{2-12} 
                                              & \multicolumn{2}{c|}{X}                              & \multicolumn{2}{c|}{Y}                              & \multicolumn{2}{c|}{S}                              & Exact                 & \multicolumn{2}{c|}{Molecular}                            & \multicolumn{2}{c|}{DNA}                                    \\ \cline{2-12} 
                                              & \multicolumn{2}{c|}{0.6}                            & \multicolumn{2}{c|}{-0.2}                           & \multicolumn{2}{c|}{0.4}                            & 0.28                  & \multicolumn{2}{c|}{0.280348}                             & \multicolumn{2}{c|}{0.280349}                               \\ \cline{2-12} 
                                              & $[X_1]$                       & $[X_0]$                       & $[Y_1]$                       & $[Y_0]$                       & $[S_1]$                       & $[S_0]$                       &                       & $[{Z_m}_1]$                       & $[{Z_m}_0]$                           & $[{Z_d}_1]$                           & $[{Z_d}_0]$                         \\ \cline{2-7} \cline{9-12} 
                                              & 0.8                      & 0.2                      & 0.4                      & 0.6                      & 0.4                      & 0.6                      &                       & 0.64                      & 0.359728                      & 0.640004                      & 0.35973                     \\ \hline
\multicolumn{1}{|l|}{\multirow{4}{*}{Exp. 2}} & \multicolumn{2}{c|}{X}                              & \multicolumn{2}{c|}{Y}                              & \multicolumn{2}{c|}{S}                              & Exact                 & \multicolumn{2}{c|}{Molecular}                            & \multicolumn{2}{c|}{DNA}                                    \\ \cline{2-12} 
\multicolumn{1}{|l|}{}                        & \multicolumn{2}{c|}{-0.6}                           & \multicolumn{2}{c|}{0.4}                            & \multicolumn{2}{c|}{0.5}                            & 0                     & \multicolumn{2}{c|}{0.00999123}                           & \multicolumn{2}{c|}{0.00998959}                             \\ \cline{2-12} 
\multicolumn{1}{|l|}{}                        & $[X_1]$                       & $[X_0]$                       & $[Y_1]$                       & $[Y_0]$                       & $[S_1]$                       & $[S_0]$                       &                       & $[{Z_m}_1]$                       & $[{Z_m}_0]$                           & $[{Z_d}_1]$                           & $[{Z_d}_1]$                         \\ \cline{2-7} \cline{9-12} 
\multicolumn{1}{|l|}{}                        & \multicolumn{1}{l|}{0.2} & \multicolumn{1}{l|}{0.8} & \multicolumn{1}{l|}{0.7} & \multicolumn{1}{l|}{0.3} & \multicolumn{1}{l|}{0.5} & \multicolumn{1}{l|}{0.5} & \multicolumn{1}{l|}{} & \multicolumn{1}{l|}{0.45} & \multicolumn{1}{l|}{0.441097} & \multicolumn{1}{l|}{0.450002} & \multicolumn{1}{l|}{0.4411} \\ \cline{1-7} \cline{9-12} 
\end{tabular}}
\label{table:muxrelu}
\end{table*}

\section{Molecular $sigmoid(2x)$}
In this section, we describe the molecular reactions for $sigmoid(2ax)$. Then the molecular $sigmoid(2ax)$ with $a=1$ can be implemented by the following molecular reactions:

\begin{eqnarray}
{X_1}_0+{A_1}_0&\rightarrow& {C_1}_1 \nonumber\\
{X_1}_0+{A_1}_1&\rightarrow& {C_1}_1 \nonumber\\
{X_1}_1+{A_1}_0&\rightarrow& {C_1}_1 \nonumber\\
{X_1}_1+{A_1}_1&\rightarrow& {C_1}_0 \nonumber\\
{A_2}_0+{C_1}_0&\rightarrow& {C_2}_0 \nonumber\\
{A_2}_0+{C_1}_1&\rightarrow& {C_2}_0 \nonumber\\
{A_2}_1+{C_1}_0&\rightarrow& {C_2}_0 \nonumber\\
{A_2}_1+{C_1}_1&\rightarrow& {C_2}_1 \nonumber\\
{X_1}_0+{C_2}_0&\rightarrow& {C_3}_1 \nonumber\\
{X_1}_0+{C_2}_1&\rightarrow& {C_3}_1 \nonumber\\
{X_1}_1+{C_2}_0&\rightarrow& {C_3}_1 \nonumber\\
{X_1}_1+{C_2}_1&\rightarrow& {C_3}_0 \nonumber\\
{A_4}_0+{C_3}_0&\rightarrow& {C_4}_0 \nonumber\\
{A_4}_0+{C_3}_1&\rightarrow& {C_4}_0 \nonumber\\
{A_4}_1+{C_3}_0&\rightarrow& {C_4}_0 \nonumber\\
{A_4}_1+{C_3}_1&\rightarrow& {C_4}_1 \nonumber\\
{X_1}_0+{C_4}_0&\rightarrow& {C_5}_1 \nonumber\\
{X_1}_0+{C_4}_1&\rightarrow& {C_5}_1 \nonumber\\
{X_1}_1+{C_4}_0&\rightarrow& {C_5}_1 \nonumber\\
{X_1}_1+{C_4}_1&\rightarrow& {C_5}_0 \nonumber\\
{A_5}_0+{C_5}_0&\rightarrow& {C_6}_0 \nonumber\\
{A_5}_0+{C_5}_1&\rightarrow& {C_6}_0 \nonumber\\
{A_5}_1+{C_5}_0&\rightarrow& {C_6}_0 \nonumber\\
{A_5}_1+{C_5}_1&\rightarrow& {C_6}_1 \nonumber\\
{X_1}_0+{C_6}_0&\rightarrow& {C_7}_1 \nonumber\\
{X_1}_0+{C_6}_1&\rightarrow& {C_7}_1 \nonumber\\
{X_1}_1+{C_6}_0&\rightarrow& {C_7}_1 \nonumber\\
{X_1}_1+{C_6}_1&\rightarrow& {C_7}_0 \nonumber\\
{X_1}_0+{C_7}_0&\rightarrow& {C_8}_1 \nonumber\\
{X_1}_0+{C_7}_1&\rightarrow& {C_8}_1 \nonumber\\
{X_1}_1+{C_7}_0&\rightarrow& {C_8}_1 \nonumber\\
{X_1}_1+{C_7}_1&\rightarrow& {C_8}_0 \nonumber\\
{C_8}_0+{C_8}_0&\rightarrow& {C_9}_1 \nonumber\\
{C_8}_0+{C_8}_1&\rightarrow& {C_9}_1 \nonumber\\
{C_8}_1+{C_8}_0&\rightarrow& {C_9}_1 \nonumber\\
{C_8}_1+{C_8}_1&\rightarrow& {C_9}_0 \nonumber\\
{C_9}_0+{C_9}_0&\rightarrow& {C_{10}}_1 \nonumber\\
{C_9}_0+{C_9}_1&\rightarrow& {C_{10}}_1 \nonumber\\
{C_9}_1+{C_9}_0&\rightarrow& {C_{10}}_1 \nonumber\\
{C_9}_1+{C_9}_1&\rightarrow& {C_{10}}_0 \nonumber\\
{E_1}_0+{C_{10}}_0&\rightarrow& W \nonumber\\
{E_1}_0+{C_{10}}_1&\rightarrow& {Z}_0 \nonumber\\
{E_1}_1+{C_{10}}_0&\rightarrow& {Z}_1 \nonumber\\
{E_1}_1+{C_{10}}_1&\rightarrow& {Z}_0+{Z}_1 \nonumber
\end{eqnarray}
where $A_1=0.2$, $A_2=0.25$, $A_3=0.3$, $A_4=0.3$, $A_5=0.5$ and $E_1=e^{-8}$ initially. $X_1$ and $Z$ represent input and output, respectively. Table~\ref{tbdata} lists computed values for $sigmoid(2x)$ at eleven equally separated points in the interval $[-1, 1]$. The table also lists the mean square error (MSE) over the eleven points.

	\begin{table*}[]
		\centering
			\caption{Computed values of $sigmoid(2x)$ with the proposed CRNs compared to exact values.}
			\vspace{1em}
			\resizebox{\textwidth}{!}{		
		\begin{tabular}{|p{59pt}|p{50pt}|c|c|c|c|c|c|c|c|c|c|c|c|}
			\hline
			\multicolumn{2}{|c|}{Function}         & $x=-1$    & $x=-0.8$  & $x=-0.6$  & $x=-0.4$  & $x=-0.2$  & $x=0$  & $x=0.2$  & $x=0.4$  & $x=0.6$  & $x=0.8$  & $x=1$    & Error                    \\ \hline
			\multirow{2}{*}{$sigmoid(2x)$}   & computed & 0.1192   & 0.1680 & 0.2315 & 0.3100 & 0.4013 & 0.5000 & 0.5988 & 0.6902 & 0.7690 & 0.8329 &  0.8822 & \multirow{2}{*}{2.77e-7} \\ \cline{2-13} %next term erro: {x^7/7!}-{x^9/9!}=2*10^-4
			& exact    & 0.1192      & 0.1680 & 0.2315 & 0.3100 & 0.4013 & 0.5 & 0.5987 & 0.6900 & 0.7685 & 0.8320 & 0.8808 &                          \\ \hline

		\end{tabular}}
            \label{tbdata}
	\end{table*}

\section{Molecular Softmax Function}
\subsection{Mass Kinetics Analysis for 3-class softmax functions}
In this section, we analyze the functionality for 3-class softmax functions. For these reactions, the mass kinetics are described by the ordinary differential equations (ODEs):

\begin{gather}
\scalebox{0.90}{$
\begin{aligned}
\frac{d[I_1]}{dt}&=[{Z_2}_0][{Z_3}_0]+[{Z_2}_0][{Z_3}_1]+[{Z_2}_1][{Z_3}_0]+[{Z_2}_1][{Z_3}_1]\nonumber\\
&-([{Z_1}_0]+[{Z_1}_1]+[{Z_2}_0]+[{Z_2}_1]+[{Z_3}_0]+[{Z_3}_1])[I_1]\nonumber \\
\frac{d[I_2]}{dt}&=[{Z_2}_0][{Z_3}_0]+[{Z_2}_0][{Z_3}_1]+[{Z_2}_1][{Z_3}_0]+[{Z_2}_1][{Z_3}_1]\nonumber\\
&-([{Z_1}_0]+[{Z_1}_1]+[{Z_2}_0]+[{Z_2}_1]+[{Z_3}_0]+[{Z_3}_1])[I_2]\nonumber \\
\frac{d[I_3]}{dt}&=[{Z_2}_0][{Z_3}_0]+[{Z_2}_0][{Z_3}_1]+[{Z_2}_1][{Z_3}_0]+[{Z_2}_1][{Z_3}_1]\nonumber\\
&-([{Z_1}_0]+[{Z_1}_1]+[{Z_2}_0]+[{Z_2}_1]+[{Z_3}_0]+[{Z_3}_1])[I_3]\nonumber \\
\frac{d[{Y_1}_1]}{dt}&=[{Z_1}_1][I_1]\nonumber\\
\frac{d[{Y_2}_1]}{dt}&=[{Z_2}_1][I_2]\nonumber\\
\frac{d[{Y_3}_1]}{dt}&=[{Z_3}_1][I_3]\nonumber\\
\frac{d[{Y_1}_0]}{dt}&=[{Z_2}_1][I_2]+[{Z_3}_1][I_3]\nonumber\\
\frac{d[{Y_2}_0]}{dt}&=[{Z_1}_1][I_1]+[{Z_3}_1][I_3]\nonumber\\
\frac{d[{Y_3}_0]}{dt}&=[{Z_1}_1][I_1]+[{Z_2}_1][I_2]\nonumber
\end{aligned}$}
\end{gather}

Assuming $[{Z_1}_0]+[{Z_1}_1]+[{Z_2}_0]+[{Z_2}_1]+[{Z_3}_0]+[{Z_3}_1]=c$, then at the equilibrium we have:

\begin{align}
[I_1] &= \frac{1}{c}([{Z_2}_0][{Z_3}_0]+[{Z_2}_0][{Z_3}_1]+[{Z_2}_1][{Z_3}_0]+[{Z_2}_1][{Z_3}_1])\nonumber \\
& = \frac{1}{c}([{Z_2}_0]+[{Z_2}_1])([{Z_3}_0]+[{Z_3}_1]) \nonumber \\
[I_2] &= \frac{1}{c}([{Z_1}_0][{Z_3}_0]+[{Z_1}_0][{Z_3}_1]+[{Z_1}_1][{Z_3}_0]+[{Z_1}_1][{Z_3}_1])\nonumber \\
& = \frac{1}{c}([{Z_1}_0]+[{Z_1}_1])([{Z_3}_0]+[{Z_3}_1]) \nonumber \\
[I_3] &= \frac{1}{c}([{Z_1}_0][{Z_2}_0]+[{Z_1}_0][{Z_2}_1]+[{Z_1}_1][{Z_2}_0]+[{Z_1}_1][{Z_2}_1])\nonumber \\
& = \frac{1}{c}([{Z_1}_0]+[{Z_1}_1])([{Z_2}_0]+[{Z_2}_1]) \nonumber \\
[{Y_1}_1] &= [{Z_1}_1][I_1] \nonumber \\
&= \frac{1}{c}[{Z_1}_1]\cdot ([{Z_2}_0]+[{Z_2}_1])([{Z_3}_0]+[{Z_3}_1]) \nonumber \\
[{Y_2}_1] &= [{Z_2}_1][I_2] \nonumber \\
&= \frac{1}{c}[{Z_2}_1]\cdot ([{Z_1}_0]+[{Z_1}_1])([{Z_3}_0]+[{Z_3}_1]) \nonumber \\
[{Y_3}_1] &= [{Z_3}_1][I_3] \nonumber \\
&= \frac{1}{c}[{Z_3}_1]\cdot ([{Z_1}_0]+[{Z_1}_1])([{Z_2}_0]+[{Z_2}_1]) \nonumber \\
[{Y_1}_0] &= \frac{1}{c}[{Z_2}_1][I_2]+\frac{1}{c}[{Z_3}_1][I_3] \nonumber \\
&= \frac{1}{c}[{Z_2}_1]\cdot ([{Z_1}_0]+[{Z_1}_1])([{Z_3}_0]+[{Z_3}_1])+\nonumber\\
&\qquad \frac{1}{c}[{Z_3}_1]\cdot ([{Z_1}_0]+[{Z_1}_1])([{Z_2}_0]+[{Z_2}_1]) \nonumber \\
[{Y_2}_0] &= \frac{1}{c}[{Z_1}_1][I_1]+\frac{1}{c}[{Z_3}_1][I_3] \nonumber \\
&= \frac{1}{c}[{Z_1}_1]\cdot ([{Z_2}_0]+[{Z_2}_1])([{Z_3}_0]+[{Z_3}_1])+\nonumber\\
&\qquad \frac{1}{c}[{Z_3}_1]\cdot ([{Z_1}_0]+[{Z_1}_1])([{Z_2}_0]+[{Z_2}_1]) \nonumber \\
[{Y_3}_0] &= \frac{1}{c}[{Z_1}_1][I_1]+\frac{1}{c}[{Z_2}_1][I_2] \nonumber \\
&= \frac{1}{c}[{Z_1}_1]\cdot ([{Z_2}_0]+[{Z_2}_1])([{Z_3}_0]+[{Z_3}_1])+\nonumber\\
&\qquad \frac{1}{c}[{Z_2}_1]\cdot ([{Z_1}_0]+[{Z_1}_1])([{Z_3}_0]+[{Z_3}_1]) \nonumber
\end{align}

The output values of the softmax function are given by:

\begin{align}
\sigma(y_1) &= \frac{[{Y_1}_1]}{[{Y_1}_1]+[{Y_1}_0]} \nonumber \\
&= \frac{\frac{[{Z_1}_1]}{[{Z_1}_1]+[{Z_1}_0]}}{\frac{[{Z_1}_1]}{[{Z_1}_1]+[{Z_1}_0]}+\frac{[{Z_2}_1]}{[{Z_2}_1]+[{Z_2}_0]}+\frac{[{Z_3}_1]}{[{Z_3}_1]+[{Z_3}_0]}} \nonumber \\
\sigma(y_2) &= \frac{[{Y_2}_1]}{[{Y_2}_1]+[{Y_2}_0]} \nonumber \\
&= \frac{\frac{[{Z_1}_1]}{[{Z_1}_1]+[{Z_1}_0]}}{\frac{[{Z_1}_1]}{[{Z_1}_1]+[{Z_1}_0]}+\frac{[{Z_2}_1]}{[{Z_2}_1]+[{Z_2}_0]}+\frac{[{Z_3}_1]}{[{Z_3}_1]+[{Z_3}_0]}} \nonumber \\
\sigma(y_3) &= \frac{[{Y_3}_1]}{[{Y_3}_1]+[{Y_3}_0]} \nonumber \\
&= \frac{\frac{[{Z_1}_1]}{[{Z_1}_1]+[{Z_1}_0]}}{\frac{[{Z_1}_1]}{[{Z_1}_1]+[{Z_1}_0]}+\frac{[{Z_2}_1]}{[{Z_2}_1]+[{Z_2}_0]}+\frac{[{Z_3}_1]}{[{Z_3}_1]+[{Z_3}_0]}} \nonumber
\end{align}

\subsection{Molecular General Softmax Functions}
In this section, we provide the molecular reactions for general softmax functions. 

Consider a standard softmax function:

\begin{eqnarray}
\sigma(y_i)&=&\frac{e^{y_i}}{\sum_{j=1}^K e^{y_j}}\nonumber\\
&=&\frac{e^{y_i}\cdot e^{-1}}{\sum_{j=1}^K (e^{y_j}\cdot e^{-1})}\nonumber\\
&=&\frac{e^{y_i-1}}{\sum_{j=1}^K e^{y_j-1}}\nonumber\\
&=&\frac{z_i}{\sum_{j=1}^K z_j}\nonumber
\end{eqnarray}
where $K$ represents the total number of different output classes and $z_j=e^{y_j-1}=\frac{[{Z_j}_1]}{[{Z_j}_1]+[{Z_j}_0]}$ for $j=1, 2, \ldots, K$. Define a set $A=\{z_1, z_2, \ldots, z_K\}$; then the softmax function $\sigma(y_i)$ can be implemented by the following molecular reactions:

For $j=1$ to $K$:

\qquad Choose one pair of $z_m$ and $z_n$, where $z_m, z_n \in A$ and $m, n \neq j$:
\begin{align}
{Z_m}_0+{Z_n}_0 &\rightarrow I_1^j  \nonumber\\
{Z_m}_0+{Z_n}_1 &\rightarrow I_1^j  \nonumber\\
{Z_m}_1+{Z_n}_0 &\rightarrow I_1^j  \nonumber\\
{Z_m}_1+{Z_n}_1 &\rightarrow I_1^j  \nonumber
\end{align}

\qquad $A = A\setminus\{z_j, z_m, z_n\}$

\qquad For $p=1$ to $K-3$:

\qquad \qquad Choose one $z_l$, where $z_l \in A$, then $A = A\setminus\{z_l\}$:
\begin{align}
I_p^j+{Z_l}_0 &\rightarrow I_{p+1}^j  \nonumber\\
I_p^j+{Z_l}_1 &\rightarrow I_{p+1}^j  \nonumber
\end{align}

For $q=1$ to $K$:
\begin{align}
{Z_q}_1+I_{K-2}^q &\rightarrow {Y_q}_1+\sum_{r=1,r\neq q}^K {Y_r}_0 \nonumber\\
{Z_q}_0+I_{K-2}^q &\rightarrow W \nonumber
\end{align}
\qquad \qquad For $r=1$ to $K$ and $r \neq q$:
\begin{align}
\qquad {Z_r}_1+I_{K-2}^q &\rightarrow W \nonumber\\
\qquad {Z_r}_0+I_{K-2}^q &\rightarrow W \nonumber
\end{align}

Notice that we use the above steps to generate the molecular reactions but simulate these reactions synchronously. Further simplifications of the above reactions are possible, but not investigated further here.

For these reactions, the mass kinetics are described by the ordinary differential equations (ODEs):

For $i=1$ to $K$:

\begin{align}
\frac{d[I_{K-2}^i]}{dt}&=\prod_{a=1, a\neq i}^K([{Z_a}_0]+[{Z_a}_1])-\frac{1}{\sum_{c=1}^K [{Z_c}_0]+[{Z_c}_1]}[I_{K-2}^i]\nonumber \\
\frac{d[{Y_i}_1]}{dt}&=[{Z_i}_1]\prod_{a=1, a\neq i}^K([{Z_a}_0]+[{Z_a}_1])\nonumber\\
\frac{d[{Y_i}_0]}{dt}&=\sum_{j=1,j\neq i}^K [{Z_j}_1]\prod_{b=1, b\neq j}^K([{Z_b}_0]+[{Z_b}_1]).\nonumber
\end{align}

Assuming $\sum_{c=1}^K [{Z_c}_0]+[{Z_c}_1]=c$, then at the equilibrium we have:

For $i=1$ to $K$: 
\begin{align}
[I_{K-2}^i]&=\frac{1}{c}\prod_{a=1, a\neq i}^K([{Z_a}_0]+[{Z_a}_1])\nonumber \\
[{Y_i}_1]&=\frac{1}{c}[{Z_i}_1]\prod_{a=1, a\neq i}^K([{Z_a}_0]+[{Z_a}_1])\nonumber\\
[{Y_i}_0]&= \frac{1}{c}\sum_{j=1,j\neq i}^K [{Z_j}_1]\prod_{b=1, b\neq j}^K([{Z_b}_0]+[{Z_b}_1]) \nonumber \\
\sigma(y_i) &= s = \frac{[{Y_i}_1]}{[{Y_i}_1]+[{Y_i}_0]} \nonumber \\
&= \frac{\frac{1}{c}[{Z_i}_1]\prod_{a=1, a\neq i}^K([{Z_a}_0]+[{Z_a}_1])}{\frac{1}{c}\sum_{j=1}^{K}[{Z_j}_1]\prod_{b=1, b\neq j}^K([{Z_b}_0]+[{Z_b}_1])} \nonumber\\
&=\frac{\frac{[{Z_i}_1]}{[{Z_i}_1]+[{Z_i}_0]}}{\sum_{j=1}^K\frac{[{Z_j}_1]}{[{Z_j}_1]+[{Z_j}_0]}}=\frac{z_i}{\sum_{j=1}^K z_j}\nonumber.
\end{align}

%% file: bare1.bbl
\begin{thebibliography}{10}

\bibitem{adleman1994molecular}
L.~M. Adleman, ``Molecular computation of solutions to combinatorial
  problems,'' {\em Science}, vol.~266, no.~5187, pp.~1021--1024, 1994.

\bibitem{samoilov2002signal}
M.~Samoilov, A.~Arkin, and J.~Ross, ``Signal processing by simple chemical
  systems,'' {\em The Journal of Physical Chemistry A}, vol.~106, no.~43,
  pp.~10205--10221, 2002.

\bibitem{thurley2014reliable}
K.~Thurley, S.~C. Tovey, G.~Moenke, V.~L. Prince, A.~Meena, A.~P. Thomas,
  A.~Skupin, C.~W. Taylor, and M.~Falcke, ``Reliable encoding of stimulus
  intensities within random sequences of intracellular ca2+ spikes,'' {\em Sci.
  Signal.}, vol.~7, no.~331, pp.~ra59--ra59, 2014.

\bibitem{sumit2015band}
M.~Sumit, R.~Neubig, S.~Takayama, and J.~Linderman, ``Band-pass processing in a
  {GPCR} signaling pathway selects for nfat transcription factor activation,''
  {\em Integrative Biology}, vol.~7, no.~11, pp.~1378--1386, 2015.

\bibitem{park2011seizure}
Y.~Park, L.~Luo, K.~K. Parhi, and T.~Netoff, ``Seizure prediction with spectral
  power of {EEG} using cost-sensitive support vector machines,'' {\em
  Epilepsia}, vol.~52, no.~10, pp.~1761--1770, 2011.

\bibitem{ghorbani2018gene}
M.~Ghorbani, E.~A. Jonckheere, and P.~Bogdan, ``Gene expression is not random:
  scaling, long-range cross-dependence, and fractal characteristics of gene
  regulatory networks,'' {\em Frontiers in physiology}, vol.~9, p.~1446, 2018.

\bibitem{parhi2019discriminative}
K.~K. Parhi and Z.~Zhang, ``Discriminative ratio of spectral power and relative
  power features derived via frequency-domain model ratio ({FDMR}) with
  application to seizure prediction,'' {\em IEEE transactions on biomedical
  circuits and systems}, 2019.

\bibitem{sauro2013synthetic}
H.~M. Sauro and K.~Kim, ``Synthetic biology: it's an analog world,'' {\em
  Nature}, vol.~497, no.~7451, p.~572, 2013.

\bibitem{sarpeshkar1998analog}
R.~Sarpeshkar, ``Analog versus digital: extrapolating from electronics to
  neurobiology,'' {\em Neural computation}, vol.~10, no.~7, pp.~1601--1638,
  1998.

\bibitem{daniel2013synthetic}
R.~Daniel, J.~R. Rubens, R.~Sarpeshkar, and T.~K. Lu, ``Synthetic analog
  computation in living cells,'' {\em Nature}, vol.~497, no.~7451, p.~619,
  2013.

\bibitem{sarpeshkar2015guest}
R.~Sarpeshkar, ``Guest editorial {-} special issue on synthetic biology,'' {\em
  IEEE transactions on biomedical circuits and systems}, vol.~9, no.~4,
  pp.~449--452, 2015.

\bibitem{teo2015synthetic}
J.~J. Teo, S.~S. Woo, and R.~Sarpeshkar, ``Synthetic biology: A unifying view
  and review using analog circuits,'' {\em IEEE transactions on biomedical
  circuits and systems}, vol.~9, no.~4, pp.~453--474, 2015.

\bibitem{soloveichik2010dna}
D.~Soloveichik, G.~Seelig, and E.~Winfree, ``{DNA} as a universal substrate for
  chemical kinetics,'' {\em Proceedings of the National Academy of Sciences},
  vol.~107, no.~12, pp.~5393--5398, 2010.

\bibitem{zhang2009control}
D.~Y. Zhang and E.~Winfree, ``Control of {DNA} strand displacement kinetics
  using toehold exchange,'' {\em Journal of the American Chemical Society},
  vol.~131, no.~47, pp.~17303--17314, 2009.

\bibitem{yurke2000dna}
B.~Yurke, A.~J. Turberfield, A.~P. Mills~Jr, F.~C. Simmel, and J.~L. Neumann,
  ``A {DNA}-fuelled molecular machine made of {DNA},'' {\em Nature}, vol.~406,
  no.~6796, p.~605, 2000.

\bibitem{turberfield2003dna}
A.~J. Turberfield, J.~Mitchell, B.~Yurke, A.~P. Mills~Jr, M.~Blakey, and F.~C.
  Simmel, ``{DNA} fuel for free-running nanomachines,'' {\em Physical review
  letters}, vol.~90, no.~11, p.~118102, 2003.

\bibitem{yurke2003using}
B.~Yurke and A.~P. Mills, ``Using {DNA} to power nanostructures,'' {\em Genetic
  Programming and Evolvable Machines}, vol.~4, no.~2, pp.~111--122, 2003.

\bibitem{gardner2000construction}
T.~S. Gardner, C.~R. Cantor, and J.~J. Collins, ``Construction of a genetic
  toggle switch in escherichia coli,'' {\em Nature}, vol.~403, no.~6767,
  p.~339, 2000.

\bibitem{weiss2003genetic}
R.~Weiss, S.~Basu, S.~Hooshangi, A.~Kalmbach, D.~Karig, R.~Mehreja, and
  I.~Netravali, ``Genetic circuit building blocks for cellular computation,
  communications, and signal processing,'' {\em Natural Computing}, vol.~2,
  no.~1, pp.~47--84, 2003.

\bibitem{jiang2013digital}
H.~Jiang, M.~D. Riedel, and K.~K. Parhi, ``Digital logic with molecular
  reactions,'' in {\em Proceedings of the International Conference on
  Computer-Aided Design}, pp.~721--727, IEEE Press, 2013.

\bibitem{jiang2011synchronous}
H.~Jiang, M.~Riedel, and K.~Parhi, ``Synchronous sequential computation with
  molecular reactions,'' in {\em Proceedings of the 48th Design Automation
  Conference}, pp.~836--841, ACM, 2011.

\bibitem{benenson2004autonomous}
Y.~Benenson, B.~Gil, U.~Ben-Dor, R.~Adar, and E.~Shapiro, ``An autonomous
  molecular computer for logical control of gene expression,'' {\em Nature},
  vol.~429, no.~6990, p.~423, 2004.

\bibitem{endy2005foundations}
D.~Endy, ``Foundations for engineering biology,'' {\em Nature}, vol.~438,
  no.~7067, p.~449, 2005.

\bibitem{ramalingam2009forward}
K.~I. Ramalingam, J.~R. Tomshine, J.~A. Maynard, and Y.~N. Kaznessis, ``Forward
  engineering of synthetic bio-logical and gates,'' {\em Biochemical
  Engineering Journal}, vol.~47, no.~1-3, pp.~38--47, 2009.

\bibitem{tamsir2011robust}
A.~Tamsir, J.~J. Tabor, and C.~A. Voigt, ``Robust multicellular computing using
  genetically encoded nor gates and chemical ‘wires’,'' {\em Nature},
  vol.~469, no.~7329, p.~212, 2011.

\bibitem{jiang2012digital}
H.~Jiang, M.~D. Riedel, and K.~K. Parhi, ``Digital signal processing with
  molecular reactions,'' {\em IEEE Design \& Test of Computers}, vol.~29,
  no.~3, pp.~21--31, 2012.

\bibitem{jiang2013discrete}
H.~Jiang, S.~A. Salehi, M.~D. Riedel, and K.~K. Parhi, ``Discrete-time signal
  processing with {DNA},'' {\em ACS synthetic biology}, vol.~2, no.~5,
  pp.~245--254, 2013.

\bibitem{salehi2015molecular}
S.~A. Salehi, H.~Jiang, M.~D. Riedel, and K.~K. Parhi, ``Molecular sensing and
  computing systems,'' {\em IEEE Transactions on Molecular, Biological and
  Multi-Scale Communications}, vol.~1, no.~3, pp.~249--264, 2015.

\bibitem{salehi2015markov}
S.~A. Salehi, M.~D. Riedel, and K.~K. Parhi, ``Markov chain computations using
  molecular reactions,'' in {\em 2015 IEEE international conference on digital
  signal processing (DSP)}, pp.~689--693, IEEE, 2015.

\bibitem{salehi2014asynchronous}
S.~A. Salehi, M.~D. Riedel, and K.~K. Parhi, ``Asynchronous discrete-time
  signal processing with molecular reactions,'' in {\em 2014 48th Asilomar
  conference on signals, systems and computers}, pp.~1767--1772, IEEE, 2014.

\bibitem{senum2011rate}
P.~Senum and M.~Riedel, ``Rate-independent constructs for chemical
  computation,'' {\em PloS one}, vol.~6, no.~6, p.~e21414, 2011.

\bibitem{kim2006construction}
J.~Kim, K.~S. White, and E.~Winfree, ``Construction of an in vitro bistable
  circuit from synthetic transcriptional switches,'' {\em Molecular systems
  biology}, vol.~2, no.~1, p.~68, 2006.

\bibitem{genot2013combinatorial}
A.~J. Genot, J.~Bath, and A.~J. Turberfield, ``Combinatorial displacement of
  {DNA} strands: application to matrix multiplication and weighted sums,'' {\em
  Angewandte Chemie International Edition}, vol.~52, no.~4, pp.~1189--1192,
  2013.

\bibitem{mjolsness1991connectionist}
E.~Mjolsness, D.~H. Sharp, and J.~Reinitz, ``A connectionist model of
  development,'' {\em Journal of Theoretical Biology}, vol.~152, no.~4,
  pp.~429--453, 1991.

\bibitem{hjelmfelt1991chemical}
A.~Hjelmfelt, E.~D. Weinberger, and J.~Ross, ``Chemical implementation of
  neural networks and turing machines,'' {\em Proceedings of the National
  Academy of Sciences}, vol.~88, no.~24, pp.~10983--10987, 1991.

\bibitem{blount2017feedforward}
D.~Blount, P.~Banda, C.~Teuscher, and D.~Stefanovic, ``Feedforward chemical
  neural network: An in silico chemical system that learns xor,'' {\em
  Artificial life}, vol.~23, no.~3, pp.~295--317, 2017.

\bibitem{mestl1996chaos}
T.~Mestl, C.~Lemay, and L.~Glass, ``Chaos in high-dimensional neural and gene
  networks,'' {\em Physica D: Nonlinear Phenomena}, vol.~98, no.~1, pp.~33--52,
  1996.

\bibitem{buchler2003schemes}
N.~E. Buchler, U.~Gerland, and T.~Hwa, ``On schemes of combinatorial
  transcription logic,'' {\em Proceedings of the National Academy of Sciences},
  vol.~100, no.~9, pp.~5136--5141, 2003.

\bibitem{hopfield1982neural}
J.~J. Hopfield, ``Neural networks and physical systems with emergent collective
  computational abilities,'' {\em Proceedings of the national academy of
  sciences}, vol.~79, no.~8, pp.~2554--2558, 1982.

\bibitem{mills1999article}
A.~P. Mills~Jr, B.~Yurke, and P.~M. Platzman, ``Article for analog vector
  algebra computation,'' {\em Biosystems}, vol.~52, no.~1-3, pp.~175--180,
  1999.

\bibitem{mills2001experimental}
A.~Mills~Jr, M.~Turberfield, A.~J. Turberfield, B.~Yurke, and P.~M. Platzman,
  ``Experimental aspects of {DNA} neural network computation,'' {\em Soft
  Computing}, vol.~5, no.~1, pp.~10--18, 2001.

\bibitem{kim2005neural}
J.~Kim, J.~Hopfield, and E.~Winfree, ``Neural network computation by in vitro
  transcriptional circuits,'' in {\em Advances in neural information processing
  systems}, pp.~681--688, 2005.

\bibitem{lim2010vitro}
H.-W. Lim, S.~H. Lee, K.-A. Yang, J.~Y. Lee, S.-I. Yoo, T.~H. Park, and B.-T.
  Zhang, ``In vitro molecular pattern classification via {DNA}-based
  weighted-sum operation,'' {\em Biosystems}, vol.~100, no.~1, pp.~1--7, 2010.

\bibitem{lopez2018molecular}
R.~Lopez, R.~Wang, and G.~Seelig, ``A molecular multi-gene classifier for
  disease diagnostics,'' {\em Nature chemistry}, vol.~10, no.~7, p.~746, 2018.

\bibitem{poole2017chemical}
W.~Poole, A.~Ortiz-Munoz, A.~Behera, N.~S. Jones, T.~E. Ouldridge, E.~Winfree,
  and M.~Gopalkrishnan, ``Chemical boltzmann machines,'' in {\em International
  Conference on DNA-Based Computers}, pp.~210--231, Springer, 2017.

\bibitem{qian2011neural}
L.~Qian, E.~Winfree, and J.~Bruck, ``Neural network computation with {DNA}
  strand displacement cascades,'' {\em Nature}, vol.~475, no.~7356, p.~368,
  2011.

\bibitem{cherry2018scaling}
K.~M. Cherry and L.~Qian, ``Scaling up molecular pattern recognition with
  {DNA}-based winner-take-all neural networks,'' {\em Nature}, vol.~559,
  no.~7714, p.~370, 2018.

\bibitem{mohammadi2017automated}
P.~Mohammadi, N.~Beerenwinkel, and Y.~Benenson, ``Automated design of synthetic
  cell classifier circuits using a two-step optimization strategy,'' {\em Cell
  systems}, vol.~4, no.~2, pp.~207--218, 2017.

\bibitem{xie2011multi}
Z.~Xie, L.~Wroblewska, L.~Prochazka, R.~Weiss, and Y.~Benenson, ``Multi-input
  rnai-based logic circuit for identification of specific cancer cells,'' {\em
  Science}, vol.~333, no.~6047, pp.~1307--1311, 2011.

\bibitem{li2015modular}
Y.~Li, Y.~Jiang, H.~Chen, W.~Liao, Z.~Li, R.~Weiss, and Z.~Xie, ``Modular
  construction of mammalian gene circuits using tale transcriptional
  repressors,'' {\em Nature chemical biology}, vol.~11, no.~3, p.~207, 2015.

\bibitem{miki2015efficient}
K.~Miki, K.~Endo, S.~Takahashi, S.~Funakoshi, I.~Takei, S.~Katayama, T.~Toyoda,
  M.~Kotaka, T.~Takaki, M.~Umeda, {\em et~al.}, ``Efficient detection and
  purification of cell populations using synthetic microrna switches,'' {\em
  Cell Stem Cell}, vol.~16, no.~6, pp.~699--711, 2015.

\bibitem{sayeg2015rationally}
M.~K. Sayeg, B.~H. Weinberg, S.~S. Cha, M.~Goodloe, W.~W. Wong, and X.~Han,
  ``Rationally designed microrna-based genetic classifiers target specific
  neurons in the brain,'' {\em ACS synthetic biology}, vol.~4, no.~7,
  pp.~788--795, 2015.

\bibitem{zhang2010dna}
D.~Y. Zhang and G.~Seelig, ``{DNA}-based fixed gain amplifiers and linear
  classifier circuits,'' in {\em International Workshop on DNA-Based
  Computers}, pp.~176--186, Springer, 2010.

\bibitem{chen2017dna}
S.~X. Chen and G.~Seelig, ``A {DNA} neural network constructed from molecular
  variable gain amplifiers,'' in {\em International Conference on DNA-Based
  Computers}, pp.~110--121, Springer, 2017.

\bibitem{salehi2018computing}
S.~A. Salehi, X.~Liu, M.~D. Riedel, and K.~K. Parhi, ``Computing mathematical
  functions using {DNA} via fractional coding,'' {\em Scientific reports},
  vol.~8, no.~1, p.~8312, 2018.

\bibitem{liu2016machine}
Y.~Liu, H.~Venkataraman, Z.~Zhang, and K.~K. Parhi, ``Machine learning
  classifiers using stochastic logic,'' in {\em 2016 IEEE 34th International
  Conference on Computer Design (ICCD)}, pp.~408--411, IEEE, 2016.

\bibitem{salehi2016chemical}
S.~A. Salehi, K.~K. Parhi, and M.~D. Riedel, ``Chemical reaction networks for
  computing polynomials,'' {\em ACS synthetic biology}, vol.~6, no.~1,
  pp.~76--83, 2016.

\bibitem{gaines1967stochastic}
B.~R. Gaines, ``Stochastic computing,'' in {\em Proceedings of the April 18-20,
  1967, spring joint computer conference}, pp.~149--156, ACM, 1967.

\bibitem{gaines1969stochastic}
B.~R. Gaines, ``Stochastic computing systems,'' in {\em Advances in information
  systems science}, pp.~37--172, Springer, 1969.

\bibitem{alaghi2013survey}
A.~Alaghi and J.~P. Hayes, ``Survey of stochastic computing,'' {\em ACM
  Transactions on Embedded computing systems (TECS)}, vol.~12, no.~2s, p.~92,
  2013.

\bibitem{brown2001stochastic}
B.~D. Brown and H.~C. Card, ``Stochastic neural computation. i. computational
  elements,'' {\em IEEE Transactions on Computers}, vol.~50, no.~9,
  pp.~891--905, 2001.

\bibitem{qian2011transforming}
W.~Qian, M.~D. Riedel, H.~Zhou, and J.~Bruck, ``Transforming probabilities with
  combinational logic,'' {\em IEEE Transactions on Computer-Aided Design of
  Integrated Circuits and Systems}, vol.~30, no.~9, pp.~1279--1292, 2011.

\bibitem{gaudet2003iterative}
V.~C. Gaudet and A.~C. Rapley, ``Iterative decoding using stochastic
  computation,'' {\em Electronics Letters}, vol.~39, no.~3, pp.~299--301, 2003.

\bibitem{naderi2011delayed}
A.~Naderi, S.~Mannor, M.~Sawan, and W.~J. Gross, ``Delayed stochastic decoding
  of {LDPC} codes,'' {\em IEEE Transactions on Signal Processing}, vol.~59,
  no.~11, pp.~5617--5626, 2011.

\bibitem{Bo2015successive}
B.~Yuan and K.~K. Parhi, ``Successive cancellation decoding of polar codes
  using stochastic computing,'' in {\em Proceedings of IEEE International
  Symposium on Circuits and Systems (ISCAS)}, pp.~3040--3043, IEEE, 2015.

\bibitem{yuan2016belief}
B.~Yuan and K.~K. Parhi, ``Belief propagation decoding of polar codes using
  stochastic computing,'' in {\em 2016 IEEE International Symposium on Circuits
  and Systems (ISCAS)}, pp.~157--160, IEEE, 2016.

\bibitem{tehrani2010majority}
S.~S. Tehrani, A.~Naderi, G.-A. Kamendje, S.~Hemati, S.~Mannor, and W.~J.
  Gross, ``Majority-based tracking forecast memories for stochastic ldpc
  decoding,'' {\em IEEE Transactions on Signal Processing}, vol.~58, no.~9,
  pp.~4883--4896, 2010.

\bibitem{liu2015lattice}
Y.~Liu and K.~K. Parhi, ``Linear-phase lattice {FIR} digital filter
  architectures using stochastic logic,'' {\em Journal of Signal Processing
  Systems}, vol.~90, no.~5, pp.~791--803, 2018.

\bibitem{KK2014architectures}
K.~K. Parhi and Y.~Liu, ``Architectures for {IIR} digital filters using
  stochastic computing,'' in {\em Proceedings of IEEE International Symposium
  on Circuits and Systems (ISCAS)}, IEEE, 2014.

\bibitem{onizawa2015gabor}
N.~Onizawa, D.~Katagiri, K.~Matsumiya, W.~J. Gross, and T.~Hanyu, ``Gabor
  filter based on stochastic computation,'' {\em IEEE Signal Processing
  Letters}, vol.~22, no.~9, pp.~1224--1228, 2015.

\bibitem{alaghi2014fast}
A.~Alaghi and J.~P. Hayes, ``Fast and accurate computation using stochastic
  circuits,'' in {\em Proceedings of the conference on Design, Automation \&
  Test in Europe}, p.~76, European Design and Automation Association, 2014.

\bibitem{qian2008synthesis}
W.~Qian and M.~D. Riedel, ``The synthesis of robust polynomial arithmetic with
  stochastic logic,'' in {\em Proceedings of the 45th annual Design Automation
  Conference}, pp.~648--653, ACM, 2008.

\bibitem{liu2017computing}
Y.~Liu and K.~K. Parhi, ``Computing polynomials using unipolar stochastic
  logic,'' {\em ACM Journal on Emerging Technologies in Computing Systems
  (JETC)}, vol.~13, no.~3, p.~42, 2017.

\bibitem{liu2016computing}
Y.~Liu and K.~K. Parhi, ``Computing hyperbolic tangent and sigmoid functions
  using stochastic logic,'' in {\em Signals, Systems and Computers, 2016 50th
  Asilomar Conference on}, pp.~1580--1585, IEEE, 2016.

\bibitem{parhi2017analysis}
K.~K. Parhi, ``Analysis of stochastic logic circuits in unipolar, bipolar and
  hybrid formats,'' in {\em 2017 IEEE International Symposium on Circuits and
  Systems (ISCAS)}, pp.~1--4, IEEE, 2017.

\bibitem{parhi2018stochastic}
K.~K. Parhi, ``Stochastic logic implementations of polynomials with all
  positive coefficients by expansion methods,'' {\em IEEE Transactions on
  Circuits and Systems II: Express Briefs}, vol.~65, no.~11, pp.~1698--1702,
  2018.

\bibitem{qian2010architecture}
W.~Qian, X.~Li, M.~D. Riedel, K.~Bazargan, and D.~J. Lilja, ``An architecture
  for fault-tolerant computation with stochastic logic,'' {\em IEEE
  transactions on computers}, vol.~60, no.~1, pp.~93--105, 2010.

\bibitem{parhi2016computing}
K.~K. {Parhi} and Y.~{Liu}, ``Computing arithmetic functions using stochastic
  logic by series expansion,'' {\em IEEE Transactions on Emerging Topics in
  Computing}, vol.~7, pp.~44--59, Jan 2019.

\bibitem{li2017neural}
B.~Li, Y.~Qin, B.~Yuan, and D.~J. Lilja, ``Neural network classifiers using
  stochastic computing with a hardware-oriented approximate activation
  function,'' in {\em 2017 IEEE International Conference on Computer Design
  (ICCD)}, pp.~97--104, IEEE, 2017.

\bibitem{chang2013architectures}
Y.-N. Chang and K.~K. Parhi, ``Architectures for digital filters using
  stochastic computing,'' in {\em Acoustics, Speech and Signal Processing
  (ICASSP), 2013 IEEE International Conference on}, pp.~2697--2701, IEEE, 2013.

\bibitem{chen2013programmable}
Y.-J. Chen, N.~Dalchau, N.~Srinivas, A.~Phillips, L.~Cardelli, D.~Soloveichik,
  and G.~Seelig, ``Programmable chemical controllers made from {DNA},'' {\em
  Nature nanotechnology}, vol.~8, no.~10, p.~755, 2013.

\bibitem{zhang2016low}
Z.~Zhang and K.~Parhi, ``Low-complexity seizure prediction from i{EEG}/s{EEG}
  using spectral power and ratios of spectral power.,'' {\em IEEE transactions
  on biomedical circuits and systems}, vol.~10, no.~3, pp.~693--706, 2016.

\bibitem{kaggle}
``American epilepsy society seizure prediction challenge.''
  http://www.kaggle.com/c/seizure-prediction.

\bibitem{liu2019computing}
X.~Liu and K.~K. Parhi, ``Computing radial basis function support vector
  machine using {DNA} via fractional coding,'' in {\em Proceedings of the 56th
  Annual Design Automation Conference 2019}, p.~143, ACM, 2019.

\bibitem{banda2013online}
P.~Banda, C.~Teuscher, and M.~R. Lakin, ``Online learning in a chemical
  perceptron,'' {\em Artificial life}, vol.~19, no.~2, pp.~195--219, 2013.

\bibitem{liu2019training}
X.~Liu and K.~K. Parhi, ``Training {DNA} perceptrons via fractional coding,''
  {\em arXiv preprint arXiv:1911.07110}, 2019.

\end{thebibliography}
